\documentclass[12pt]{article}

\usepackage{bm}
\usepackage{float}
\usepackage{relsize}
\usepackage{array}
\usepackage{fullpage}
\usepackage{amsfonts}
\usepackage{amsmath}
\usepackage{setspace}
\usepackage{easybmat}
\usepackage{slashed}
\usepackage{amsmath}
\usepackage{amssymb}
\usepackage{graphicx}
\usepackage{verbatim}
\usepackage{caption}
\usepackage{subcaption}
\usepackage{makecell}
\usepackage{hyperref}

\usepackage{hyperref}

\usepackage{environ}
\usepackage{tikz}

\usetikzlibrary{decorations.markings}
\usetikzlibrary{arrows.meta}
\usetikzlibrary{decorations.pathreplacing}

\allowdisplaybreaks
\def\timesbox{\hbox{$\scriptscriptstyle\times$}}
\def\ant{ {{\lower 1ex  \timesbox} \atop {\raise 1.5ex  \timesbox}}}
\def\f{\frac}
\def\pa{\partial}
\def\non{\nonumber\\}

\def\b{\beta}

\def\d{\delta}
\def\e{\epsilon}
\def\g{\gamma}

\def\l{\lambda}
\def\m{\mu}
\def\n{\nu}

\def\p{\pi}

\def\s{\sigma}
\def\t{\tau}
\def\th{\theta}
\def\D{\Delta}

\def\G{\Gamma}
\def\L{\Lambda}

\newcommand{\mathsym}[1]{{}}
\newcommand{\unicode}[1]{{}}

\newcommand{\be}{\begin{equation}}
\newcommand{\ee}{\end{equation}}
\newcommand{\beqa}{\begin{eqnarray}}
\newcommand{\eeqa}{\end{eqnarray}}
\newcommand{\bsp}{\begin{split}}
\newcommand{\esp}{\end{split}}
\newcommand{\bgth}{\begin{gather}}
\newcommand{\egth}{\end{gather}}

\newcommand{\tr}{\hbox{tr}}

\title{Neumann scalars in AdS: partition functions and phases }
\author{Biki Bishwakarma$^{a}$\footnote{bikibishwakarma8759@gmail.com}, Astha Kakkar$^{b}$\footnote{asthakakkar8@gmail.com}   and 
Swarnendu Sarkar$^{a}$\footnote{swarnen@gmail.com}\\
$^a$\small{{\em Department of Physics, Vidyasagar University,}}\\
\small{{\em Midnapore 721102, India}}\\
$^b$\small{{\em Indian Institute of Technology Kanpur,
}}\\ 
\small{{\em Kalyanpur, Kanpur 208016, India}}\\
}

\date{}

\begin{document}

\maketitle
\abstract{We present an analysis of one-loop partition functions for scalars in AdS$_{d+1}$ obeying the Neumann boundary condition and explore phases of scalar field theories in several dimensions at zero and finite temperature. The partition function computation involves an analytic continuation from $\D_+$ corresponding to the Dirichlet boundary condition to $\D_-$ corresponding to Neumann boundary condition. We show that this can be implemented by deformations of the contour for integral over eigenvalue ($\l$) of the Laplace operator as compared to the integral over ${\mathbb R}$ in \cite{Kakkar:2022hub} for the Dirichlet boundary condition. We further contrast these phases with those appearing for the case of scalars obeying the Dirichlet boundary condition and corroborate the occurrence of these phases by studying the long range behaviour of correlators. }

\newpage

\tableofcontents

\section{Introduction}
Quantum field theories in Anti de-Sitter (AdS) spaces offer a possible path to get an insight into the flat space theories, by taking the large AdS radius limit, which otherwise often face difficulties such as infrared divergences when studied directly on flat space \cite{Callan:1989em}. Apart from this, in AdS space, these studies have often led to interesting results that are not visible in flat space theories or go away when one takes the flat space limit \cite{Inami:1985dj}. Both these facts have served as motivations for studying field theories on AdS spaces \cite{Burgess:1984ti} - \cite{Sugishita:2016iel} in early years and extending over to relatively recent years in \cite{Carmi:2018qzm} - \cite{Antunes:2025iaw}, including AdS RG flows \cite{Meineri:2023mps} - \cite{Antunes:2024hrt}.  

In all these studies the boundary conditions on the fields play a crucial role. It has long been known \cite{Breitenlohner:1982jf} that for a stable AdS vacuum a free scalar field of mass $m$ in AdS$_{d+1}$, with $m^2 >-d^2/4+1$, only Dirichlet boundary condition is allowed, while for $-d^2/4< m^2< -d^2/4+1$ both Dirichlet and Neumann boundary conditions are possible. The lower bound $m^2=-d^2/4$ is famously known as the Breitenlohner-Freedman (BF) bound. That, these constraints can alternatively be seen using the AdS/CFT correspondence \cite{Aharony:1999ti} is now widely known. A scalar field solution near the conformal boundary in Poincar\'e coordinates (\ref{poincare}) behaves as  \cite{Klebanov:1999tb}
\beqa\label{kwsly}
\phi(\vec{x},y)\sim y^{\Delta_-}\left[\phi_0(\vec{x})+O(y^2)\right]+ y^{\Delta_+}\left[A(\vec{x})+O(y^2)\right]
\eeqa
where $\Delta_{\pm}=d/2\pm\sqrt{d^2/4+m^2}$. The term containing $A(\vec{x})$ is normalizable and corresponds to physical excitations in AdS which fall off at the conformal boundary at $y\rightarrow0$. The term containing $\phi_0(\vec{x})$ is non-normalizable and is interpreted as the source for the boundary scalar operator $\mathcal{O}(\vec{x})$ dual to the field $\phi(\vec{x},y)$ in AdS. While Dirichlet boundary conditions correspond to the standard quantization of the bulk field, Neumann boundary conditions lead to an alternate quantization in which the roles of the source and expectation value in the dual conformal field theory are interchanged. They thus correspond to different dual CFT on the boundary. The fact that, for unitarity, the conformal dimensions $\Delta_{\pm}$ must be real and non-negative such that $\Delta_{\pm} \ge (d-2)/2$ leads to the above constraints on $m^2$.  

While studies on quantum field theories in AdS have mostly focused on Dirichlet boundary condition, newer studies with Neumann boundary condition include studying quantum field theories, phases and confinement \cite{Copetti:2023sya}, \cite{Ciccone:2024guw}, \cite{DiPietro:2025ozw}, \cite{Ciccone:2025dqx}, \cite{Ankur:2026ylr}, boundary RG flows and $F$-theorems \cite{Giombi:2020rmc}, \cite{Giombi:2021cnr}, \cite{Nishioka:2021uef}, \cite{Bason:2025sxb}, \cite{Bason:2025zpy} and explorations of gravity with a new boundary term \cite{Krishnan:2016mcj}, \cite{Krishnan:2016dgy}.

A somewhat recent area of exploration has been the extension of field theory studies to thermal AdS, leading to the study of effective potentials and phases at non zero temperatures. To the leading order, the one-loop partition functions in AdS with conformal boundary $S^1 \times S^{d-1}$ take quite simple forms. The procedure to evaluate these partition functions for fields in AdS obeying Dirichlet boundary condition is well established in the literature \cite{Gibbons:2006ij}-\cite{Kakkar:2026nqp}. In one of such methods discussed in \cite{Kakkar:2022hub}-\cite{Kakkar:2026nqp} the authors have derived the partition functions for scalars, spinors and $U(1)$ vectors by solving the eigenvalue problem for the corresponding Laplace operators for AdS. In the adoption of the same procedure for the Neumann boundary condition one faces hindrances in a direct computation of the one-loop partition function, $Z^{(1)}$ as in this case the bulk to bulk propagator does not define a square-integrable function in the invariant volume measure for AdS \cite{Carmi:2018qzm}. One thus analytically continues the result obtained for the Dirichlet boundary condition in terms of $\Delta_{+}$ to that for the Neumann boundary condition with $\D_-$ effectively replacing $\Delta_{+}$ with $\Delta_{-}$.
An alternate route discussed in \cite{Carmi:2018qzm} leading to the derivation of the partition function involves deforming the contour for the integral over the spectral parameter.

In this paper we discuss two other such possible deformations which lead to the same result for Neumann boundary condition and obtain the one loop determinants $Z^{(1)}$ for scalars from the well established ones for the Dirichlet boundary condition. This is realized by deforming the contour for the eigenvalue $\l$ (of the Laplace operator) integral in the evaluation presented in \cite{Kakkar:2022hub}. This equates to an analytical continuation from $\D_+$ to $\D_-$ in the expressions for $Z^{(1)}$ as compared to the results in \cite{Kakkar:2022hub}.

The major portion of the paper is devoted to the study of phases of scalar field theories with Neumann boundary conditions. 
We begin by studying the simpler single scalar theory with $\phi^4$ interaction, then the models with $O(N)$ symmetry both at finite and large $N$. We study these theories both at zero and non-zero temperatures. The phases at zero temperature are studied in terms of the mass-squared $m^2$ and the AdS radius $L$ and at finite temperature on the $\beta-m^2$ plane. The Neumann boundary condition imposes more stringent conditions on the effective masses of theories which leads to quite different phase plots as compared to the Dirichlet boundary condition. In some cases the functional form of the effective potential prevents access to the symmetry broken phase. Another interesting outcome for the large $N$ models in all the dimensions we considered is that the constraint imposed by the requirement that the finite temperature contribution to the effective potential must be convergent rules out a further chunk of the phase space otherwise allowed by the unitarity bounds. In this case the requirement that $\b(d/2-\nu)>0$ prevents us from accessing the $M^2=0$ solutions and hence prevents the occurrence of symmetry breaking. The results are summarized in Table \ref{tab1}. In the table various phases are mentioned as unstable as the potential is strictly concave downwards. These plots are bounded by the restriction on $\D_-$. The solution to the saddle point equations though unstable at $\phi_{cl}=0$, $\phi_{cl}$ does not roll down infinitely below but settles down at the end points of these plots (see for example Figure \ref{ads4ssmL}).
In some cases where results for Dirichlet boundary condition are not already available in \cite{Kakkar:2022hub} - \cite{Kakkar:2026nqp} we give these here for comparison. We also verify the occurrences of the various phases by studying the long range behaviour of the correlators for these theories.

\begin{table}[H]
\centering
\renewcommand{\arraystretch}{1.4}
\begin{tabular}{|c|c|c|c|c|}
\hline
Dimension & Model &  Zero Temperature Phases & Finite Temperature Phases\\
\hline
 & Single Scalar & SB & SP \&  SB\\
AdS$_2$ & $O(N)$ & SB (vanishes as $N$ increases) & SP \&  SB  \\
 & Large N & SP (unstable) & SP (unstable) \\
\hline
 & Single Scalar & SP \& SB & SP \& SB  \\
AdS$_3$ & $O(N)$ &  SP \& SB & SP \&  SB  \\
 & Large N & SB & SP (unstable) \\ 
 \hline
 & Single Scalar & SP (unstable) & SP (unstable)  \\
AdS$_4$ & $O(N)$ & SP (unstable) & SP (unstable)  \\
 & Large N & SP (unstable) & SP (unstable) \\
\hline
\end{tabular}
\caption{Summary of phases of QFTs in AdS with Neumann boundary condition across various dimensions. Here SB stands for symmetry breaking and SP for symmetry preserving.}
\label{tab1}
\end{table}

The outline of the rest of the paper is as follows: In section \ref{effaction} we lay down the method and the contour prescriptions used to evaluate effective potentials in AdS for scalars obeying Neumann boundary condition at zero and then at finite temperature. In section \ref{phases} we discuss the phases of the single scalar, $O(N)$ model and large $N$ model numerically for $d=1,2,3$ and provide relevant expressions for $d=4$; and then further provide an analytical justification for these by studying the behaviour of two-point correlators. In section \ref{discussion} we finally conclude and provide a plan for further study of field theories in AdS with Neumann boundary condition. Appendix \ref{thermaltrace} gives the details of the trace computations for thermal AdS while appendix \ref{ads4corr} collects some lengthy expressions used in the main text.

\section{One-loop effective action at zero and finite temperature} \label{effaction}
In \cite{Kakkar:2022hub} we studied effective potentials and one-loop partition functions for scalar field theories with the Dirichlet boundary condition. In this section we extend the analysis to the case of scalar field theories with Neumann boundary condition.
The effective potential is written as
\begin{equation}
\label{c1e1}
V^{}_{eff}(\phi^{}_{cl})=-\frac{1}{{\cal V}^{}_{d+1}}\log Z=\frac{1}{2{\cal V}^{}_{d+1}}\mbox{tr} \log[-\square^{}_{E}+V''(\phi^{}_{cl})]+V(\phi^{}_{cl}).
\end{equation}
where

\begin{eqnarray}
Z&=&\det[-\square^{}_{E}+V''(\phi^{}_{cl})]^{-1/2} \exp(-{\cal V}^{}_{d+1} V(\phi^{}_{cl}))\non 
&=&\exp\left(-\frac{1}{2}\mbox{tr} \log[-\square^{}_{E}+V''(\phi^{}_{cl})]\right)\exp\left(-{\cal V}_{d+1} V(\phi^{}_{cl})\right)
\end{eqnarray}

where we have expanded about the constant classical value of $\phi$, which extremizes the potential, as $\phi=\phi^{}_{cl}+\eta$ and integrated and have used

\begin{equation}
\square^{}_{E}=\frac{\partial^{}_{\mu}[\sqrt{g}g^{\mu\nu} \partial^{}_{\nu}]}{\sqrt{g}}.
\end{equation}
 ${\cal V}^{}_{d+1}$ here is the volume of $d+1$ dimensional Euclidean space.
Extremizing the effective potential
\begin{equation}
\label{c1e2}
\frac{dV^{}_{eff}}{d\phi^{}_{cl}}=V'(\phi^{}_{cl})+\frac{1}{2{\cal V}^{}_{d+1}}\mbox{tr}\left[ \frac{1}{-\square^{}_{E}+V''(\phi^{}_{cl})}\right] V'''(\phi^{}_{cl}).
\end{equation}

We now lay down the computation of the trace in (\ref{c1e2}) for Euclidean AdS$_{d+1}$ with Neumann boundary condition. The metric of AdS$_{d+1}$ in Poincar\'e coordinates is
\beqa\label{poincare}
ds^2=\f{L^2}{y^2}\left(dy^2+\eta_{\mu\nu}dx^{\mu}dx^{\nu}\right).
\eeqa

In these coordinates we first solve the following eigenvalue equation,

\beqa\label{eeqn}
-L^2 \Box_E \Psi_{\l,\vec{k}}(\vec{x},y)=\left[\l^2+\left(\frac{d}{2}\right)^2\right] \Psi_{\vec{k},\l}(\vec{x},y)
\eeqa

where,

\beqa\label{box}
-L^2 \Box_E=-y^2\left[y^{d-1}\pa_y\left(y^{1-d}\pa_y\right)+\pa_{\vec{x}}^2\right].
\eeqa

 The solution to this equation gives the eigenfunctions of (\ref{eeqn}) and are given by ${\phi}_{\l}(ky)=(ky)^{d/2}K_{i\l}(ky)$ and $\Psi_{\vec{k},\l}(\vec{x},y)=\phi_\l(k y)e^{\pm i \vec{k}.\vec{x}}$.

The trace in (\ref{c1e2}) can be computed using the spectral form by closely following \cite{Kakkar:2022hub} as given below

\beqa\label{trace1}
&&\frac{1}{L^2}\tr\left[\frac{1}{- \Box_E+V^{''}(\phi_{cl})}\right]\non
&=&\f{1}{L^{2}}\int d^{d+1}x \sqrt{g}\int d\l ~\m(\l)\int \frac{d^d k}{(2\pi)^d}\f{1}{(L^{d+1}k^d)}\langle\l,k|\left[\frac{1}{- \Box_E+V^{''}(\phi_{cl})}\right]|y,\vec{x}\rangle\langle y,\vec{x}|\l,k\rangle\non
&=&\f{1}{L^{d+1}}\int d^{d+1}x \sqrt{g}\int  \f{d\l~\m(\l)}{\l^2 +\n^2}\int \frac{d^d k}{(2\pi)^d} y^d K^2_{i\l}(ky)
\eeqa

where $\n=\sqrt{\left(\f{d}{2}\right)^2 + L^2 V^{''}(\phi_{cl})}$ and 

\beqa
\m(\l)=\f{2 \l}{\p^2}\sinh(\p\l)
\eeqa

Now, the trace can be computed by performing the integrals in various orders. One can first integrate over $k$ in (\ref{trace1}) and further perform the integral over $\l$ by closing the contour in the upper half of the complex $\l$ plane. Or we may use identities of special functions to compute the trace following another order of integrations. These computations corresponding to the Dirichlet boundary condition give the trace as

\beqa\label{trace2}\frac{1}{L^2}\tr\left[\frac{1}{- \Box_E+V^{''}(\phi_{cl})}\right]=\f{{\cal V}_{d+1}}{L^{d+1}}\frac{\Gamma(\Delta) \Gamma(1/2-d/2)}{\Gamma(1-d+\Delta)(4\pi)^{(d+1)/2}}
\eeqa

with $\Delta=\Delta_{+}=d/2+\nu$. For Neumann boundary condition, we can analytically continue the expression for the trace from $\Delta_{+}$ to $\Delta_{-}$. In the following we give a ``derivation" through contour prescription of the $\lambda$ integral in (\ref{trace1}).
\subsection{Euclidean AdS and contour prescription}

In this section we discuss two contour prescriptions leading to the same answer for the zero temperature trace corresponding to the Neumann boundary condition.

\subsubsection{Prescription 1}
We begin with the expression for the Dirichlet case
\begin{eqnarray}\label{eq2.12}
\frac{1}{L^{2}}\mbox{tr} \left(\frac{1}{-\square^{}_{E}+V''(\phi^{}_{cl})}\right)=\frac{1}{L^{d+1}}\int d^{d+1}x \sqrt{g} \int_0^{\infty} \frac{d\lambda \mu(\lambda)}{\lambda^{2}+\n^{2}} \int \frac{d^{d}(ky)}{(2\pi)^{d}} [K^{}_{i\lambda}(ky)]^{2}
\end{eqnarray}

Integrating over $k$ and plugging in the normalization measure then gives 

\begin{equation}
\frac{1}{L^{2}}\mbox{tr} \left(\frac{1}{-\square^{}_{E}+V''(\phi^{}_{cl})}\right)=\frac{{\cal V}^{}_{d+1}}{L^{d+1}2^{d+1} \pi^{\frac{d+1}{2}} \Gamma(\frac{d}{2}+\frac{1}{2})}\int_{-\infty}^{\infty}\frac{d\lambda}{\lambda^{2}+\n^{2}}  \frac{\Gamma(\frac{d}{2} \pm i\lambda)}{\Gamma(\pm i\lambda)}
\end{equation}
where $\Gamma(\pm a)=\Gamma(a)\Gamma(-a)$.\\

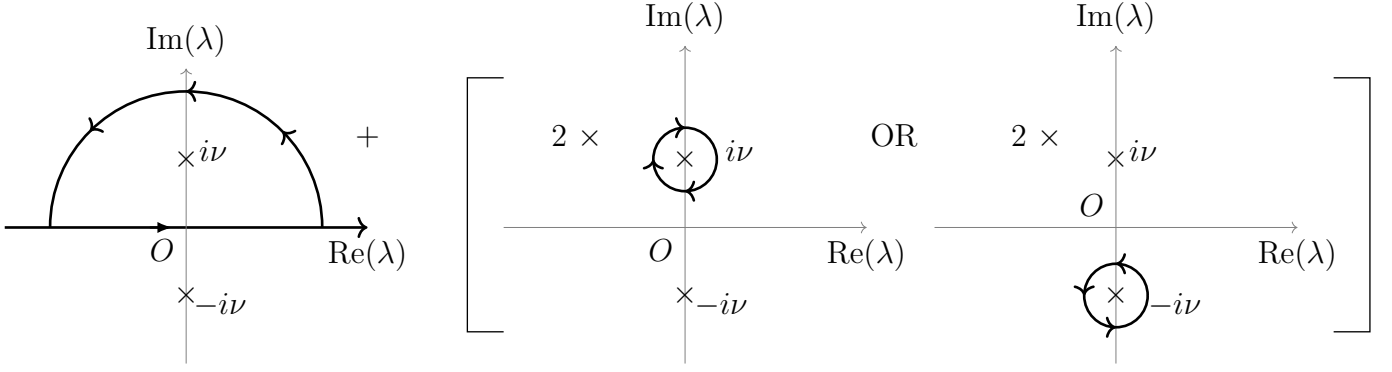
\begin{figure}[H]
\begin{tikzpicture}
  [scale=0.6, line width=1pt,
    decoration={%
      markings,
      mark=at position 0.25 with {\arrow[line width=1pt]{>}},
      mark=at position 0.5 with {\arrow[line width=1pt]{>}},    
      mark=at position 0.75 with {\arrow[line width=1pt]{>}}     
    }   
  ]
  \begin{scope}[xshift=0cm]
  
  \draw [->] (-4,0) -- (4,0) coordinate (xaxis);
  \draw [help lines,->] (0,-3)coordinate(-yaxis) -- (0,3.5) coordinate (yaxis);
  \node  at (0,1.5) {$\times$};
    \node  at (0.6,1.7) {$i\nu$};
 \draw[-{Latex[length=2mm]},thick]
(-0.8,0) -- (-0.3,0);
  \draw [postaction=decorate] (3,0) arc (0:180:3) ;
 \node [below] at (xaxis) {$\text{Re}(\l)$};
  \node [above] at (yaxis) {$\text{Im}(\l)$};
  \node [below left] {$O$};
  \node  at (0,-1.5) {$\times$};
    \node  at (0.75,-1.7) {$-i\nu$};
    \node at (4,2) {\text{+} };
  \end{scope}

  \begin{scope}[xshift=11cm]
  \draw [help lines,->] (-4,0) -- (4,0) coordinate (xaxis);
  \draw [help lines,->] (0,-3)coordinate(-yaxis) -- (0,4) coordinate (yaxis);
  \node  at (0,1.5) {$\times$};
    \node  at (1.2,1.7) {$i\nu$};
(-0.8,0) -- (-0.3,0);
  \draw [ postaction=decorate] (0.7,1.5) arc (360:0:0.7) ;
  \node [below] at (xaxis) {$\text{Re}(\l)$};
  \node [above] at (yaxis) {$\text{Im}(\l)$};
  \node [below left] {$O$};
  \node  at (0,-1.5) {$\times$};
    \node  at (0.8,-1.7) {$-i\nu$};
     \node at (-2.5,2) {\text{~2} $\times$};
  \end{scope}

  \begin{scope}[xshift=20.5cm]
  \node  at (0,1.5) {$\times$};
    \node  at (0.6,1.7) {$i\nu$};
  \draw [help lines,->] (-4,0) -- (4,0) coordinate (xaxis);
  \draw [help lines,->] (0,-3)coordinate(-yaxis) -- (0,4) coordinate (yaxis);
  \node  at (0,-1.5) {$\times$};
    \node  at (1.3,-1.7) {$-i\nu$};
(-0.8,0) -- (-0.3,0);
  \draw [postaction=decorate] (0.7,-1.5) arc (0:360:0.7) ;
  \node [below] at (xaxis) {$\text{Re}(\l)$};
  \node [above] at (yaxis) {$\text{Im}(\l)$};
  \node [above left] {$O$};
   \node at (-3.3,2) {\text{OR ~~~~~~~~2} $\times$};
  \end{scope}
  \draw[line width=0.5pt]
 (7.0,-2.3) -- (6.2,-2.3)
  -- (6.2,3.3)
  -- (7.0,3.3);
  \draw[line width=0.5pt]
 (25.3,3.3) -- (26.1,3.3)
  -- (26.1,-2.3)
  -- (25.3,-2.3);
\end{tikzpicture}
 \caption{contour 1}
  \label{contour1}
 \end{figure}

Next the $\lambda$ integral needs to be performed. For Dirichlet boundary condition the integral over $\lambda$ is performed over the real line $\mathbb{R}$. For the Neumann case we deform the contour of the integral as shown in Figure \ref{contour1}. The poles for $\lambda$ occur when
\begin{enumerate}
    \item $\lambda^{2}+\n^{2}=0$
    \item $\frac{d}{2}\pm i\lambda=-n$
\end{enumerate}
The first contour in the UHP in Figure \ref{contour1} gives the same contribution as in Dirichlet boundary condition. For the Neumann boundary condition we combine contributions from the two contours in Figure \ref{contour1}. This computation is shown below. \\

\noindent 
(1) For
\begin{equation}
\lambda^{2}+\n^{2}=0
\end{equation}
we have $\lambda=\pm i\n$. We close the contour in the upper half so that $\lambda=+ i\n$ and the residue from this pole is 
\begin{eqnarray}
\mbox{Residue}
=-\Gamma\left(\frac{d}{2}\pm \n\right) \sin(\pi \n)
\end{eqnarray}
\noindent (2) The Gamma functions also give poles when
\begin{equation}
\frac{d}{2}\pm i\lambda=-n
\end{equation}
for $n=0,1,2, \cdots$.
However only $\Gamma(\frac{d}{2}+i\lambda)$ has poles in the upper half. Therefore
\begin{equation}
(2 \pi i)\mbox{Residue}=(2 \pi i)\left[\sum^{\infty}_{n=0} \frac{1}{\n^{2}-\left(\frac{d}{2}+n\right)^{2}}\frac{(-1)^{n}}{n!} \Gamma(d+n)i \left(n+\frac{d}{2}\right) \frac{\sin(\pi(n+\frac{d}{2}))}{\pi}\right].\end{equation}
This simplifies to
\begin{equation}
\frac{(-2) \pi^{2} \cos(\pi \n) \tan(\frac{d\pi}{2})}{(\cos(\pi d)-\cos(2\pi \n))\Gamma((1-(\frac{d}{2}+\n))\Gamma(1-(\frac{d}{2}-\n))}.
\end{equation}
After combining the two contributions while taking either of the two combinations shown in Figure \ref{contour1} we get

\begin{equation}
\label{em1}
\frac{1}{L^{2}}\mbox{tr} \left(\frac{1}{-\square^{}_{E}+V''(\phi^{}_{cl})}\right)=\frac{{\cal V}^{}_{d+1}}{4(\pi)^{\frac{d+2}{2}}} \frac{\Gamma(\frac{d}{2})}{\Gamma(d)}\frac{1}{L^{d+1}}\frac{\Gamma(\frac{d}{2}\pm \n) \sin(\pi(\frac{d}{2}+\n))}{\cos(\frac{\pi d}{2})}.
\end{equation}

which is the result (\ref{trace2}) with $\Delta=\Delta_{-}$.

\vspace{1cm}

\subsubsection{Prescription 2}
Another representation of the trace can obtained from the integral in equation (\ref{eq2.12}) by using the following identities \cite{Grad}, \cite{Watson} (see also \cite{Miyagawa:2015sql}):

\beqa\label{ikrel}
K_{i\l}(ky)=\f{\p}{2}\f{I_{-i\l}(ky)-I_{i\l}(ky)}{i\sinh(\p\l)}~,~~I_{\pm i\l}\left(\f{yy^{\prime}}{2s}\right)=\frac{1}{2\p i}\int_{\infty-i\p}^{\infty+i\p} dt \exp\left(\frac{yy^{\prime}}{2s}\cosh (t)\mp i\l t\right)
\eeqa
and for $y< y^{\prime}$,

\beqa
I_{-i\l}(ky)K_{i\l}(ky^{\prime})&=& \int_0^{\infty} \f{ds}{2s}e^{-k^2 s} e^{-\f{y^2+y^{\prime 2}}{4s}}I_{-i\l}\left(\f{yy^{\prime}}{2s}\right)\non
&=& \int_0^{\infty} \f{ds}{2s}e^{-k^2 s} e^{-\f{y^2+y^{\prime 2}}{4s}} \frac{1}{2\p i}\int_{\infty-i\p}^{\infty+i\p} dt \exp\left(\frac{yy^{\prime}}{2s}\cosh (t)+i\l t\right)
\eeqa

thus we can write

\beqa
\frac{1}{L^2}\tr\left[\frac{1}{- \Box_E+V^{''}(\phi_{cl})}\right]&=& \f{1}{L^{d+1}}\int d^{d+1}x \sqrt{g} ~ y^d  \int \frac{d^d k}{(2\pi)^d}\f{1}{(2\pi i)}\int_{-\infty}^{\infty} d \l \frac{2\l}{\l^2+\n^2}\\
&\times& \left[\lim_{y^{\prime}\rightarrow y}\int\f{ds}{2s}e^{-k^2 s} e^{-\f{y^2+y^{\prime 2}}{4s}} \frac{1}{2\p i}\int_{\infty-i\p}^{\infty+i\p} dt \exp\left(\frac{yy^{\prime}}{2s}\cosh (t)+i\l t\right)\right].\nonumber
\eeqa

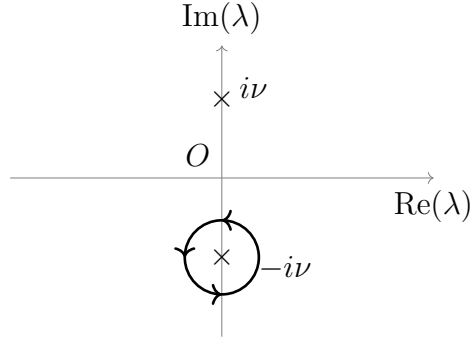
\begin{figure}[H]
\begin{center}
\begin{tikzpicture}
  [scale=0.7, line width=1pt,
    decoration={%
      markings,
      mark=at position 0.25 with {\arrow[line width=1pt]{>}},
      mark=at position 0.5 with {\arrow[line width=1pt]{>}},    
      mark=at position 0.75 with {\arrow[line width=1pt]{>}}     
    }   
  ]

  \draw [help lines,->] (-4,0) -- (4,0) coordinate (xaxis);
  \draw [help lines,->] (0,-3)coordinate(-yaxis) -- (0,2.5) coordinate (yaxis);
  \node  at (0,-1.5) {$\times$};
    \node  at (1.2,-1.7) {$-i\nu$};
(-0.8,0) -- (-0.3,0);
  \draw [postaction=decorate] (0.7,-1.5) arc (0:360:0.7) ;
  \node [below] at (xaxis) {Re$(\l)$};
  \node [above] at (yaxis) {Im$(\l)$};
  \node [above left] {$O$};
   \node  at (0,1.5) {$\times$};
    \node  at (0.6,1.7) {$i\nu$};

\end{tikzpicture}
\end{center}
\caption{Contour 2}
\label{contour2}
\end{figure}

This is the integral representation of the trace in which $\lambda$ integral is performed in UHP for Dirichlet boundary condition. For the Neumann boundary condition we will perform the contour integral over $\lambda$ over the contour as shown in Figure \ref{contour2}. This leads to the following expression for the trace

\beqa \label{con2}
&&\f{1}{L^{d+1}}\int d^{d+1}x \sqrt{g} ~ y^d  \int \frac{d^d k}{(2\pi)^d}\left[\lim_{y^{\prime}\rightarrow y}\int\f{ds}{2s}e^{-k^2 s} e^{-\f{y^{2}+y^{\prime 2}}{4s}}\frac{1}{2\p i}\int_{\infty-i\p}^{\infty+i\p} dt \exp\left(\frac{yy^{\prime}}{2s}\cosh (t)+\n t\right)\right]\non\label{order3}\\
&=&\f{1}{L^{d+1}}\int d^{d+1}x \sqrt{g} ~ y^d  \int \frac{d^d k}{(2\pi)^d}\left[\lim_{y^{\prime}\rightarrow y}\int \f{ds}{2s}e^{-k^2 s} e^{-\f{y^{2}+y^{\prime 2}}{4s}}I_{-\n}\left(\f{yy^{\prime}}{2s}\right)\right]\label{order2}\\
&=& \f{1}{L^{d+1}}\int d^{d+1}x \sqrt{g} ~ y^d  \int \frac{d^d k}{(2\pi)^d} \left[\lim_{y^{\prime}\rightarrow y}I_{-\n}(ky)K_{\n}(ky^{\prime})\right]\label{ikform} \\
&=&\f{{\cal V}_{d+1}}{L^{d+1}}\f{\G\left(d/2-\n\right)\G(1/2-d/2)}{\G\left(1-d/2-\n\right) (4\p)^{(d+1)/2}}\label{trzt}
\eeqa

The above expression has been checked to match with the result in (\ref{em1}). Equation (\ref{trzt}) gives the analytically continued result valid for $d \ge 1$. The final form given by (\ref{ikform}) will be used in computation of the trace on thermal AdS using contour 2 in Figure \ref{contour2}.

\subsection{Thermal AdS}

Using the contour prescription shown in Figure \ref{contour2} we give the results for thermal AdS below. To simplify we put $L=1$ and put back factors of $L$ when required. The details of the computations which are essentially adaptation of the computation in \cite{Kakkar:2022hub} are given in appendix \ref{thermaltrace}. 
Thermal AdS$_3$ is defined as the quotient space $\mathbb{H}^3/\mathbb{Z}$ with the metric 
\beqa
ds^2=\f{1}{y^2}\left(dy^2+dzd\bar{z}\right)
\eeqa
and with the following action of $\g^{n}\in \mathbb{Z}$ on the coordinates
\beqa
\g^n(y,z)=(e^{-n\beta}y, e^{2\pi in\tau}z) ~~~~\mbox{where}~~~~\tau=\frac{1}{2\pi}(\theta+i\beta)
\eeqa
This can be generalized for higher dimensions as is discussed in appendix \ref{thermaltrace}.
The trace for AdS$_{3}$ can thus be written as

 \beqa
\tr\left[\frac{1}{- \Box_E+V^{''}(\phi_{cl})}\right]
= \frac{2}{{\cal N}}\sum_{n=1}^{\infty}\int \f{dy}{y}\f{e^{-n\beta}}{|1-e^{2\pi i n\t}|^2} \f{e^{\b n\n}}{(-2\n)}
\eeqa

which gives

\beqa\label{finalp3n}
\log Z^{(1)}_{\t}=\sum_{n=1}^{\infty}\f{1}{n}\f{e^{-n\beta(1-\n)}}{|1-e^{2\pi i n\t}|^2} 
\eeqa

Similarly for AdS$_{d+1}$ for even and odd $d$ we respectively have

\beqa\label{zd1}
\log Z^{(1)}_{\t}=\sum_{n=1}^{\infty}\f{e^{n\beta\n}}{n}\prod_{i=1}^{d/2}\f{e^{-n\beta}}{|1-e^{2\pi i n\t_i}|^2} 
\eeqa

\beqa\label{zdto1}
\log Z^{(1)}_{\t}=\sum_{n=1}^{\infty}\f{e^{-n\beta(1/2-\n)}}{n|1-e^{-n\beta}|}\prod_{i=1}^{(d-1)/2}\f{e^{-n\beta}}{|1-e^{2\pi i n\t_i}|^2} 
\eeqa

\section{Phases} \label{phases}
In this section we study the phases of scalar theories in AdS both at zero and finite temperature.  We consider the theories with a single scalar field and then the $O(N)$ vector model for finite $N$ and finally in the large $N$ limit and decipher some interesting results. We use the following regularized volumes in our computations: $\mathcal{V}_2=-\b L,  \mathcal{V}_3 = -\pi \b L^2/2, \mathcal{V}_4 = 2 \pi \b L^3/3, \mathcal{V}_5 = 3 \pi^2 \b L^4/16 $ for the finite temperature analysis. We give elaborate analysis of the potentials for $d=1,2,3$ but we restrict ourselves to only providing expressions for $d=4$.
\subsection{Single Scalar}

We begin with the following Lagrangian for the $\phi^{4}$ theory
\begin{equation}
\label{ch4e1}
\mathcal{L}_E=\frac{1}{2} ( \partial ^{}_{\mu} \phi)^{2} + \frac{1}{2} m^{2} \phi^2 + \frac{\lambda}{4!}  \phi ^{4}.
\end{equation}
The effective potential for this theory at finite temperature can thus be written as

\beqa
V^{}_{eff}(\phi^{}_{cl})&=&V(\phi_{cl})-\f{1}{{\cal V}_{d+1}}[\log Z^{(1)}+\log Z^{(1)}_{\t}].
\eeqa

\subsubsection{AdS$_2$}\label{ads2singles}
We now consider the theory on AdS$_2$. The zero temperature trace (\ref{trzt}) is divergent for $d=1$. We thus expand 

\beqa\label{expads2}
\f{1}{{2\cal V}_{d+1}L^2}\mbox{tr} \frac{1}{-\square_E+M^{2}}=\frac{1}{4 \pi L^2 \e}+\frac{1}{8\pi L^2}\left[-2 \psi^{(0)}\left(\n+\frac{1}{2}\right)-\gamma +\log (4\pi )+2\pi\tan(\pi \nu) \right]
\eeqa
where $\e=1-d$, $\gamma$ is the Euler-Mascheroni constant and $\psi^{(m)}(x)=d^{m+1} \log\Gamma(x)/dx^{m+1}$ the Polygamma function. In the above expression the divergence appears as a pole $1/\e$. 
The renormalized effective potential is obtained by including a counterterm 
\begin{eqnarray}\label{ads2effpot}
V^{}_{eff}(\phi^{}_{cl})=\frac{1}{2} m^{2}\phi^{2}_{cl}+\frac{\lambda}{4!}\phi^{4}_{cl} +M^2L^2\d m^2-\f{1}{{\cal V}_{d+1}}(\log Z^{(1)}+\log Z^{(1)}_{\t}).
\end{eqnarray}
The counterterm $\d m^2$ is obtained by imposing the following renormalization condition on the zero temperature effective potential $V_{eff}^0$
\beqa\label{sads2rc}
\left.\f{\pa^2}{\pa \phi_{cl}^2}V^{0}_{eff}(\phi^{}_{cl})\right|_{\phi_{cl}=0}=m^2
\eeqa
where from (\ref{ads2effpot}) and after reintroducing the factors of $L$, $V^{0}_{eff}$ is written as

\beqa
V^{0}_{eff}=\frac{1}{2} m^{2}\phi^{2}_{cl}+\frac{\lambda}{4!}\phi^{4}_{cl} +M^2 L^2\d m^2-\f{1}{2{\cal V}_{d+1} L^2 }\int_{M^2 L^2}^{\infty}\tr \left[\frac{1}{-\square_E+M^2}\right]d(ML)^2.
\eeqa
With the effective mass $M^2=m^2+\f{\l}{2}\phi_{cl}^2$ , this gives

\beqa
\d m^2&=&-\f{1}{2{\cal V}_{d+1} L^2}\tr \left[\frac{1}{-\square_E+M^2 }\right]_{\phi_{cl}=0}\\
&=&-\f{1}{4\pi L^2\e}-\f{1}{8\pi L^2}\left[-2 \psi^{(0)}\left(\n+\frac{1}{2}\right)-\g+\log(4\pi) +\log (4\pi)+ 2\pi\tan\left(\pi \nu \right)\right]_{\phi_{cl}=0}\nonumber
\eeqa
where $\nu=\sqrt{\f{1}{4}+M^2L^2}$. Further, the integral over $M^2$ is written as

\beqa
=-\frac{1}{4\pi L^2}\int^{M^{2}L^2}_{0} d(M^2 L^2) \left[\psi^{(0)}\left(\n+\frac{1}{2}\right)-\pi\tan\left(\pi \nu \right)\right]~~+~~\mbox{infinite constant}.
\eeqa
We shall discard this infinite constant in our analysis. Next including the finite temperature contribution from (\ref{zdto1}) and putting $\theta=0, {\cal V}^{}_{2}=-\beta L$ , the effective potential at finite temperature can be written as 

\beqa\label{ads2n}
V^{}_{eff}(\phi^{}_{cl})&=&\frac{1}{2} m^{2}\phi^{2}_{cl}+\frac{\lambda}{4!}\phi^{4}_{cl}
+\frac{1}{4\pi L^2}\int^{M^{2}L^2}_{0} d x^2 \left[\psi^{(0)}\left(\sqrt{\f{1}{4}+m^2L^2}+\frac{1}{2}\right)\right.\nonumber \\
&-&\left.\psi^{(0)}\left(\sqrt{\f{1}{4}+x^2}+\frac{1}{2}\right)+\pi \tan\left(\pi \sqrt{\f{1}{4}+x^2}\right)-\pi \tan\left(\pi\sqrt{\f{1}{4}+m^2L^2} \right) \right]\nonumber\\
&+&\f{1 }{\f{\b}{L} L^2}\sum^{\infty}_{n=1}\frac{1}{n}\frac{e^{-n\f{\beta}{L}\left(\frac{1}{2}-\nu(M^2 L^2)\right)}}{\left|1-e^{-n\f{\beta}{L}}\right|}
\eeqa

The potential (\ref{ads2n}) must satisfy the following bounds:

\begin{enumerate}
    \item BF bound: $M^2 L^2 \ge -1/4$ so that the argument under the square-root is positive. 
    \item Unitarity bound (for $d=1$): $\Delta_-\geq 0$ which implies, $M^2 L^2 \le 0$
\end{enumerate}

\begin{figure}[t] 
\begin{center} 
  \begin{minipage}{0.4\textwidth}%
   \begin{subfigure}[b]{0.9\linewidth}
    \centering
    \includegraphics[width=1\linewidth]{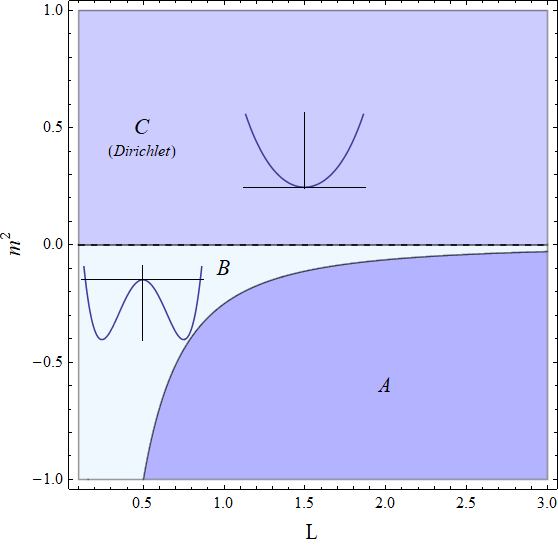} 
    \caption{} 
    \label{ads2ssmL} 
    \vspace{1ex}
  \end{subfigure}
  \end{minipage}
 \begin{minipage}{0.4\textwidth}%
   \begin{subfigure}[b]{0.95\linewidth}
    \centering
    \includegraphics[width=1\linewidth]{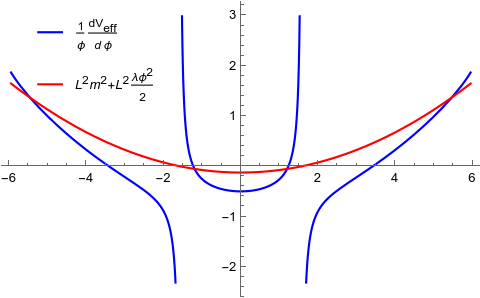} 
    \caption{} 
    \label{valid} 
    \vspace{1ex}
  \end{subfigure}
\end{minipage}
  \caption{ AdS$_2$ single scalar: (a) Zero temperature $m^2-L$ phase plot for Neumann (Dirichlet) boundary condition with $\lambda=0.4$ (b) Shows a point $m^2=-0.5, L=0.5, \l=0.4$ for Neumann boundary condition when the $\phi_{cl}\ne 0$ roots satisfy the unitarity bound. } 
  \label{DNAds2} 
  \end{center} 
\end{figure}

\noindent
{\bf Zero temperature:}\\
The derivative of the potential (\ref{ads2n}) at zero temperature is

\beqa\label{ads2nprime}
\frac{1}{\phi_{cl}}\frac{\partial V(\phi_{cl})}{{\partial \phi_{cl}}}&=&m^{2}+\frac{\lambda}{6}\phi^{2}_{cl}
+\frac{\l}{4\pi} \left[\psi^{(0)}\left(\sqrt{\f{1}{4}+m^2L^2}+\frac{1}{2}\right)-\psi^{(0)}\left(\sqrt{\f{1}{4}+M^2L^2}+\frac{1}{2}\right)\right.\nonumber \\
&+&\left.\pi \tan\left(\pi \sqrt{\f{1}{4}+M^2L^2}\right)-\pi \tan\left(\pi\sqrt{\f{1}{2}+m^2L^2} \right) \right]
\eeqa

For Dirichlet boundary condition the unitarity bound is always satisfied and the only constraint is the BF bound. The corresponding phase diagram is shown in Figure (\ref{ads2ssmL}).

Figure \ref{valid} shows a plot of $\frac{1}{\phi_{cl}}\frac{\partial V(\phi_{cl})}{{\partial \phi_{cl}}}$ and $M^2 L^2$ versus $\phi_{cl}$ for Neumann boundary condition. The derivative of the potential in this case has infinite number of disconnected branches with corresponding zeros coming from the $\tan$ function. Only the central curve satisfies the unitarity bound i.e roots of the saddle point equation are such that $M^2 L^2 \le 0$.
For $m^2<0$ the curve $M^2 L^2$ is convex and always have real roots. When $M^2 L^2\rightarrow 0$, $\frac{1}{\phi_{cl}}\frac{\partial V(\phi_{cl})}{{\partial \phi_{cl}}} \rightarrow \infty$ because of the first $\tan$ term in equation (\ref{ads2nprime}) and thus the equation is not satisfied. Thus whenever the central curve of the derivative of the potential has a root, it always satisfies the unitarity bound. In deriving the phase plot only the BF bound plays the constraining role. The phase plot along with representative potential plots is thus similar to the the one for the Dirichlet case as shown in Figure \ref{ads2ssmL} for regions $A$ and $B$ while a distinction between the two appears in the region $C$ which is not allowed for the Neumann boundary condition.  The boundary separating the regions $A-B$ is given by the curve $m^2L^2=-1/4$. As the term under the square-roots  $\sqrt{\f{1}{4}+m^2L^2}$ in (\ref{ads2nprime}) becomes negative when $m^2L^2<-1/4$ region $A$ does not have a potential plot.

\begin{figure}[H] 
\begin{center} 
  \begin{minipage}{0.5\textwidth}%
   \begin{subfigure}[b]{0.8\linewidth}
    \centering
    \includegraphics[width=1.1\linewidth]{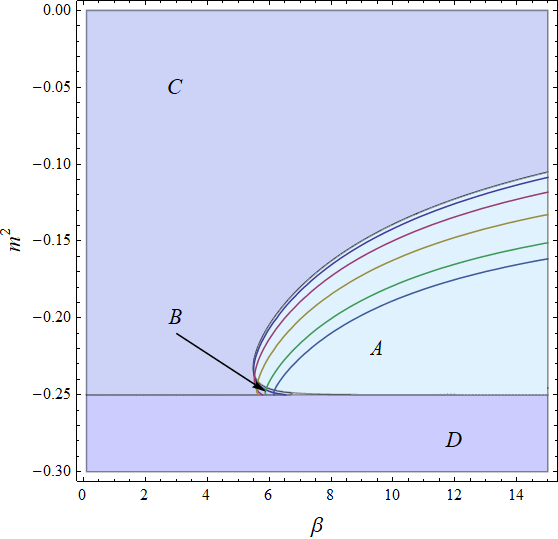} 
    \caption{Phases on the $\b-m^2$ plane} 
    \label{ads2ssmT} 
    \vspace{1ex}
  \end{subfigure}
  \end{minipage}
  \begin{minipage}{0.5\textwidth}%
  \begin{subfigure}[b]{0.5\linewidth}
    \centering
    \includegraphics[width=0.95\linewidth]{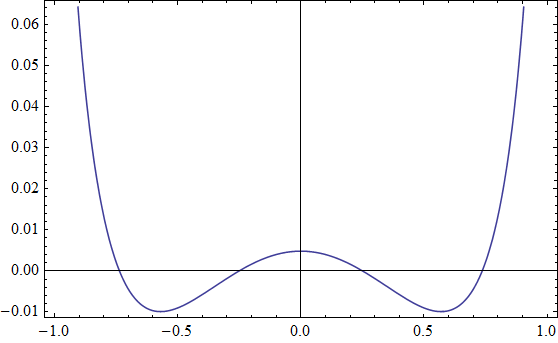} 
    \caption*{$A$: $m^2=-0.22$, $\b=9.5$} 
    \vspace{1ex}
  \end{subfigure}%
  \begin{subfigure}[b]{0.5\linewidth}
    \centering
    \includegraphics[width=0.95\linewidth]{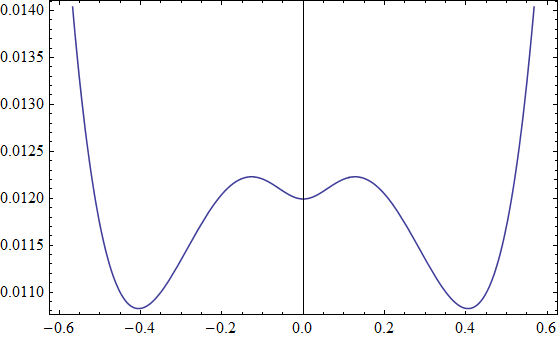} 
    \caption*{$B1$: $m^2=-0.248$, $\b=5.9$} 
    \vspace{1ex}
  \end{subfigure} 
  \begin{subfigure}[b]{0.5\linewidth}
    \centering
    \includegraphics[width=0.95\linewidth]{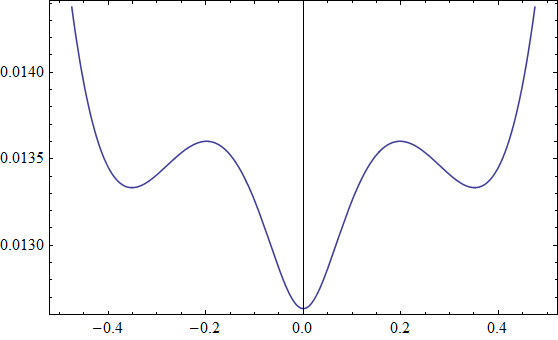} 
    \caption*{$B2$: $m^2=-0.249$, $\b=5.7$} 
  \end{subfigure}
  \begin{subfigure}[b]{0.5\linewidth}
    \centering
    \includegraphics[width=0.95\linewidth]{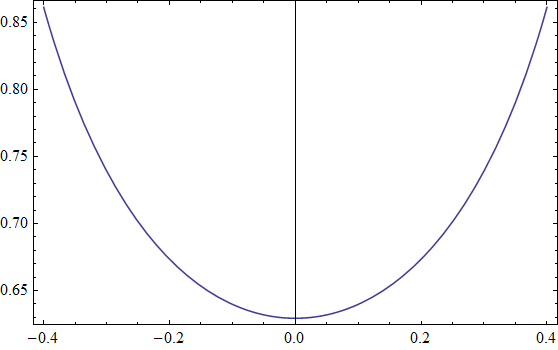} 
   \caption*{$C$: $m^2=-0.05$, $\b=3$} 
  \end{subfigure} 
\end{minipage}
  \caption{AdS$_2$ single scalar: Phases and potentials for Neumann boundary condition and $L=0.4$, $\l=0.4$ and $n=10$. The potential plots are shown for regions $A$, $B$ and $C$.}
  \label{ads2ss1} 
  \end{center} 
\end{figure}

\noindent
{\bf Finite temperature:}\\
The finite temperature phase plot with potentials are shown in Figure \ref{ads2ss1}. Figure \ref{ads2ssmT} shows contours with extrema at various values of $\phi_{cl}$. The intersections of these contours implies the existence of additional extrema. This is the region $B$ in the phase plot. As one move inside the region from region $C$ new minima (other than $\phi_{cl}=0$) appear, which then exchange dominance as the temperature is lowered and finally ends up in a double well potential in the region $A$.  Representative potential plots corresponding to regions $A-C$ are shown alongside the phase plot in Figure \ref{ads2ssmT}. \\

\begin{figure}[H] 
\begin{center} 
  \begin{minipage}{0.45\textwidth}%
   \begin{subfigure}[b]{0.9\linewidth}
    \centering
    \includegraphics[width=1\linewidth]{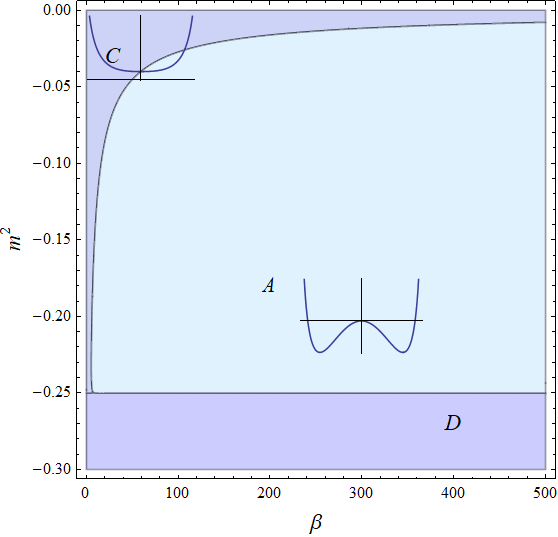} 
    \caption{} 
    \label{ads2ssmT1} 
    \vspace{1ex}
  \end{subfigure}
  \end{minipage}
 \begin{minipage}{0.45\textwidth}%
   \begin{subfigure}[b]{0.9\linewidth}
    \centering
    \includegraphics[width=1\linewidth]{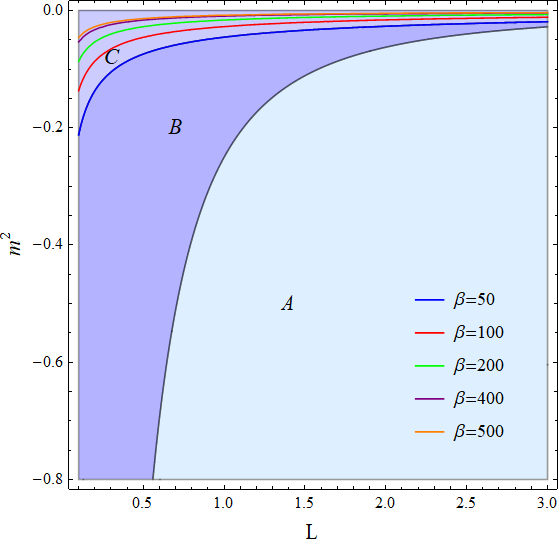} 
    \caption{} 
    \label{ads2ssmL1} 
    \vspace{1ex}
  \end{subfigure}
\end{minipage}
  \caption{AdS$_2$ single scalar: (a) The $m^2$ vs $\b$ phase plot for Neumann boundary condition asymptotes to the zero temperature case for high $\b$ and $\l=0.4, L=1$  (b) The $m^2$ vs $L$ phase plot for Neumann boundary condition for various values of $\b$ and $\l=0.4$} 
  \end{center} 
\end{figure}

Figure \ref{ads2ssmT1} shows the $m^2-\beta$ phase plot for low temperatures (high $\beta$ values). Figure \ref{ads2ssmL1} shows the $m^2-L$ phase plot for various values of $\beta$. It may be noted that the boundary separating the regions $B-C$ shifts with decreasing $\beta$ thereby shrinking the region $B$ of the phase plot (as compared to that in Figure \ref{ads2ssmL}) as the temperature is increased. At zero temperature the $B-C$ boundary, across which the second order transition occurs, approaches the horizontal line at $m^2=0$. The $A-B$ boundary however is unaffected by $\beta$ as it is defined by the BF bound $m^2L^2=-1/4$. 

\subsubsection{AdS$_3$}\label{ads3singles}

We shall now consider the theory on AdS$_3$. Thus setting $d=2$ in (\ref{trzt}) gives

\beqa\label{trzt3}
\f{1}{2 L^2}\mbox{tr}\left[\f{1}{-\square_E+V^{''}(\phi_{cl})}\right]=\f{{\cal V}_3}{L^3}\f{\sqrt{1+M^2 L^2}}{8\pi}
\eeqa
Note that while the magnitude of the expression is the same for both the Dirichlet and Neumann boundary conditions the overall sign of the trace is different.  Here $\n$ and the effective mass, $M$ are related as

\begin{equation}
\label{ch4e3}
\n=\sqrt{1+M^2 L^2}=\left[1+L^2\left(\f{\lambda }{2}\phi^{2}_{cl}+m^{2}\right)\right]^{1/2}.
\end{equation}

$\log Z^{(1)}$ is obtained from (\ref{trzt3}) after integrating over $M^2$ as in equation (\ref{intads3n}). $\log Z^{(1)}_{\t}$ is  given in equation (\ref{finalp3n}). 
We further set the angular potential $\theta=0$ and insert ${\cal V}_{3}=-\beta \pi L^2/2$. 

The complete expression for the one-loop corrected effective potential at zero temperature for the Dirichlet and Neuman boundary conditions are,

\begin{eqnarray}\label{ads3finitet1}
V^{(0)}_{eff}(\phi^{}_{cl})=\frac{1}{2} m^{2}\phi^{2}_{cl}+\frac{\lambda}{4!}\phi^{4}_{cl} \mp \frac{\n^3}{12\pi L^3}
\end{eqnarray}

while at finite temperature we get
\begin{eqnarray}\label{ads3finitet1t}
V^{(\t)}_{eff}(\phi^{}_{cl})=\frac{1}{2} m^{2}\phi^{2}_{cl}+\frac{\lambda}{4!}\phi^{4}_{cl} \mp \frac{\n^3}{12\pi L^3} +\f{2}{\pi  L^3  \f{\b}{L}}\sum _{n=1}^{\infty} \frac{ e^{-\f{\b}{L} n (1\pm \n)}}{ n \left(1-e^{-\f{\b}{L} n}\right)^2}
\end{eqnarray}
\noindent

The upper and the lower signs in the above expressions are for Dirichlet and Neumann boundary conditions respectively.\\

\noindent
\textbf{Zero Temperature:}\\
We first analyze the zero temperature part of the potential. Since the expression does not involve any transcendental function the analysis of phases can be done analytically. We first note that BF bound reads

\beqa\label{BF}
M^2 L^2=L^2\left(m^2+\f{\l}{2}\phi_{cl}^2 \right) \geq -1
\eeqa

For an extremum to exist at $\phi_{cl}=0$ this bound requires $m^2 L^2 >-1$. Apart from the BF bound another constraint comes from the  
unitarity bound, which is

\beqa\label{UB}
\D&\geq& \f{d-2}{2}=0\nonumber\\
\D_{\pm}&=&\f{d}{2}\pm\sqrt{\left(\f{d}{2}\right)^2+M^2L^2}\ge 0
\eeqa

For Dirichlet boundary condition ($\D=\D_+$) this bound is always satisfied. However for the Neumann boundary condition, the unitarity bound gives a nontrivial constraint

\beqa\label{UBN}
\D_{-}\geq 0 ~\rightarrow ~ M^2 L^2 ~\leq~ 0.
\eeqa
For non zero values of $\phi_{cl}$ this implies
\beqa\phi_{cl}^2\leq-\f{2 m^2}{\l}.
\eeqa

We now look at the phase plots constrained by the above bounds. The derivative of the potential is
\beqa
\f{d V(\phi_{cl})}{d \phi_{cl}}&=&\phi_{cl}\left(m^2+\f{\l}{6}\phi_{cl}^2 \mp \f{\l}{8\pi L}\sqrt{1+L^2\left(m^2+\f{\l}{2}\phi_{cl}^2\right)} \right).
\eeqa

For a $\phi_{cl}\ne 0$ solution to exist we have

\beqa \label{ads3sol}
m^2+\f{\l}{6}\phi_{cl}^2 \mp \f{\l}{8\pi L}\sqrt{1+L^2\left(m^2+\f{\l}{2}\phi_{cl}^2\right)}=0
\eeqa
and we get the upper bound by setting $\phi_{cl}=0$. This gives
\beqa \label{ads3ssupb}
m^2 \mp \f{\l}{8\pi L}\sqrt{1+m^2 L^2}=0.
\eeqa

\begin{figure}[H] 
\begin{center} 
  \begin{minipage}{0.45\textwidth}%
   \begin{subfigure}[b]{0.9\linewidth}
    \centering
    \includegraphics[width=1\linewidth]{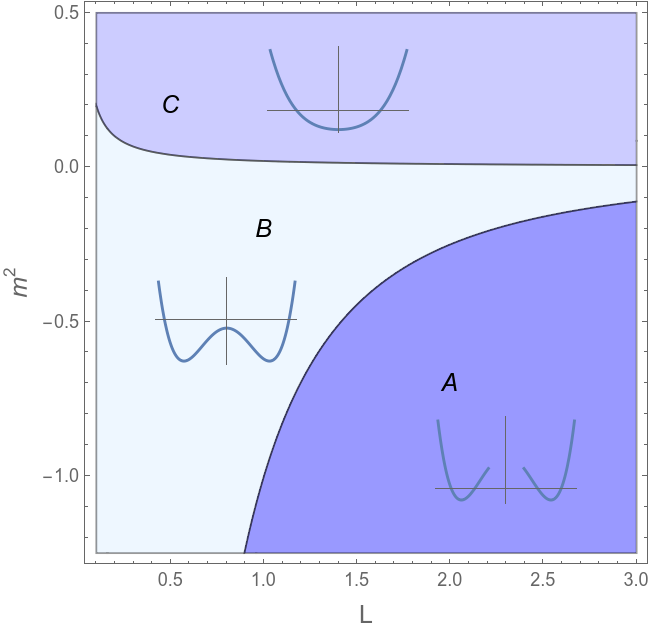} 
    \caption{} 
    \label{ads3ssd} 
    \vspace{1ex}
  \end{subfigure}
  \end{minipage}
 \begin{minipage}{0.45\textwidth}%
   \begin{subfigure}[b]{0.9\linewidth}
    \centering
    \includegraphics[width=1\linewidth]{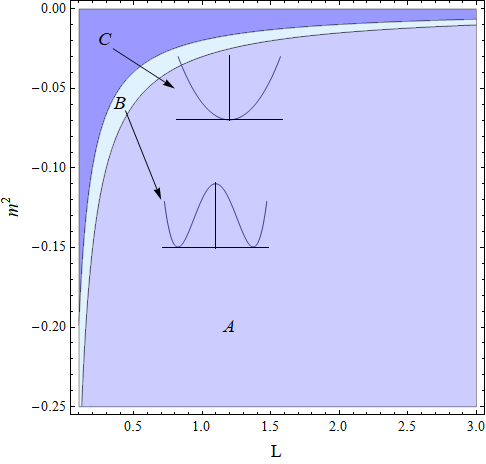} 
    \caption{} 
    \label{ads3ssn} 
    \vspace{1ex}
  \end{subfigure}
\end{minipage}
  \caption{AdS$_3$ single scalar: Zero temperature $m^2-L$ phase plot for (a) Dirichlet and (b) Neumann boundary conditions with $\lambda=0.5$} 
  \label{DNAds3} 
  \end{center} 
\end{figure}

For the Dirichlet case, the above equation with the minus sign gives the upper bound for the $\phi_{cl}\ne 0$ solution. The lower bound where all three extrema namely at $\phi_{cl}= 0$ and $\phi_{cl}\ne 0$ exist is given by the BF bound  (\ref{BF}). On the $L-m^2$ plane below this bound although the  $\phi_{cl}\ne 0$ solution exists $\phi_{cl}= 0$ solution does not exist. The potential plot is thus truncated around $\phi_{cl}= 0$. Figure \ref{ads3ssd} shows the phase plot showing these bounds and potentials. The boundary between $A-B$ is given by the BF bound while $B-C$ represents the second order phase transition.

For the Neumann boundary condition the equation with the lower (plus) sign in (\ref{ads3sol}) gives the upper bound for $\phi_{cl}\ne 0$ solution to exist. Further since we need our solutions from (\ref{ads3sol}) to satisfy the unitarity bound (\ref{UBN}) we get another lower bound at
\beqa \label{ads3ssub}
m^2 L=-\f{3 \l}{16 \pi}.
\eeqa

Figure \ref{ads3ssn} shows these bounds and potentials for the Neumann boundary condition. Here the boundary between $A-B$ is the unitary bound given by equation (\ref{ads3ssub}) while $B-C$ gives the upper bound for the symmetry breaking phase \ref{ads3ssupb}. The BF bound lies within region $A$.

\noindent
\textbf{Finite Temperature:}\\
At finite temperature we get a phase plot with features and potential plots similar to AdS$_2$ Figure \ref{ads2ss1}.

\subsubsection{AdS$_4$}\label{ads4singles}
Expanding the expression for the trace at zero temperature around $d = 3$ 
\beqa\label{trzt4}
\f{1}{{2\cal V}_{d+1} L^2}\mbox{tr} \frac{1}{-\square_E+M^{2}}&=&\f{(M^2 L^2+2)}{32 \pi^2 L^4}\left[-\f{2}{\epsilon}+\g-1-\log 4\pi+\psi^{(0)}\left(\f{3}{2}+\nu \right)+\psi^{(0)}\left(-\f{1}{2}+\nu \right) \right.\nonumber\\
&-& 2 \pi \tan(\pi \nu)\Bigg]
\eeqa
To renormalize we set the following renormalization conditions at zero temperature,

\beqa
\left.\f{\partial V_{eff}}{\partial \phi_{cl}^2}\right|_{\phi_{cl}=0}=\f{m^2}{2} ~~~~,~~~~ \left.\f{\partial^2 V_{eff}}{\partial^2 \phi_{cl}^2}\right|_{\phi_{cl}=0}=\f{\l}{12}
\eeqa
which give us

\beqa
\d m^2=-\f{\l}{2 {\cal V}_{d+1}}\mbox{tr} \left[\frac{1}{-\square+M^{2}}\right]_{\phi_{cl}=0}~~~\mbox{and}~~~\f{\d \l}{6}=-\f{\l^2}{{\cal V}_{d+1}}\f{\partial}{\partial \phi_{cl}^2}\mbox{tr} \left[\frac{1}{-\square+M^{2}}\right]_{\phi_{cl}=0}.
\eeqa
This gives us after including the finite temperature contribution the following effective potential

\beqa
V_{eff}&=&\f{1}{2}m^2 \phi_{cl}^2+\f{1}{4!}\l\phi_{cl}^4+\int_0^{\phi_{cl}^2}\f{\l(2+M^2 L^2)}{32 \pi^2 L^2}\left[ \psi^{(0)}\left(\f{3}{2}+\nu(M^2L^2)\right)+\psi^{(0)}\left(-\f{1}{2}+\nu(M^2L^2)\right)\right.\nonumber\\
&-&\left.\psi^{(0)}\left(\f{3}{2}+\nu(m^2L^2)\right)-\psi^{(0)}\left(-\f{1}{2}+\nu(m^2L^2)\right)-2\pi \tan\left(\pi \nu(M^2L^2)\right)+2\pi\tan(\pi \nu(m^2L^2)) \right] d \phi_{cl}^2\nonumber\\
&-&\f{\l^2(2+m^2 L^2)}{256 \pi^2 ~\nu(m^2L^2)}\left(  \psi^{(1)}\left(\f{3}{2}+\nu(m^2L^2)\right)+\psi^{(1)}\left(-\f{1}{2}+\nu(m^2L^2)\right)-2\pi^2\sec^2\left(\pi \nu (m^2 L^2)\right)\right)\phi_{cl}^4 \nonumber\\
&-&\sum _{n=1}^{\infty} \frac{3 e^{-n\f{\b}{L}\left(\f{3}{2}-\nu(M^2 L^2) \right)}}{2 L^4 \f{\b}{L} \pi \left|1-e^{-n\f{\b}{L}} \right|^3}.
\eeqa

The derivative of the potential at zero temperature is

\beqa\label{ads4sprime}
\frac{1}{\phi_{cl}}\frac{\partial V(\phi_{cl})}{{\partial \phi_{cl}}}&=&m^{2}+\frac{\lambda}{6}\phi^{2}_{cl}+\f{\l(2+M^2 L^2)}{16 \pi^2 L^2}\left[ \psi^{(0)}\left(\f{3}{2}+\nu(M^2L^2)\right)+\psi^{(0)}\left(-\f{1}{2}+\nu(M^2L^2)\right)\right.\nonumber\\
&-&\left.\psi^{(0)}\left(\f{3}{2}+\nu(m^2L^2)\right)-\psi^{(0)}\left(-\f{1}{2}+\nu(m^2L^2)\right)-2\pi \tan\left(\pi \nu(M^2L^2)\right)\right. \nonumber\\
&+&2\pi\tan\left(\pi \nu(m^2L^2)\right) \Big]-\f{\l^2(2+m^2 L^2)}{64 \pi^2 ~\nu(m^2L^2)}\Bigg(  \psi^{(1)}\left(\f{3}{2}+\nu(m^2L^2)\right)\nonumber\\
&+&\psi^{(1)}\left(-\f{1}{2}+\nu(m^2L^2)\right)-2\pi^2\sec^2\left(\pi \nu (m^2 L^2)\right)\Bigg)\phi_{cl}^2. 
\eeqa
The unitarity bound for the Neumann boundary condition now requires
\beqa
\D_-\geq \f{d-2}{2}=\f{1}{2} \rightarrow M^2 L^2 \leq -\f{5}{4}
\eeqa
 as now
 \beqa
 \D_\pm=\f{3}{2}\pm\sqrt{\f{9}{4}+M^2L^2}
 \eeqa
 while the BF bound is $M^2 L^2 \geq -9/4$.
 
\begin{figure}[H] 
\begin{center} 
  \begin{minipage}{0.45\textwidth}%
   \begin{subfigure}[b]{0.9\linewidth}
    \centering
    \includegraphics[width=1\linewidth]{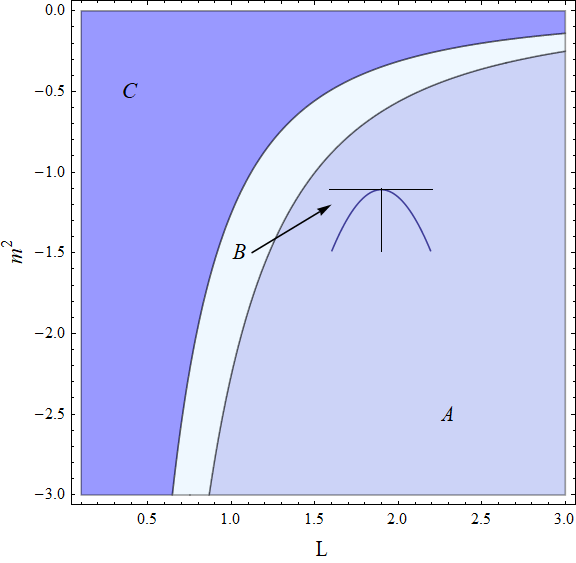} 
    \caption{} 
    \label{ads4ssmL} 
    \vspace{1ex}
  \end{subfigure}
  \end{minipage}
 \begin{minipage}{0.45\textwidth}%
   \begin{subfigure}[b]{0.9\linewidth}
    \centering
    \includegraphics[width=1\linewidth]{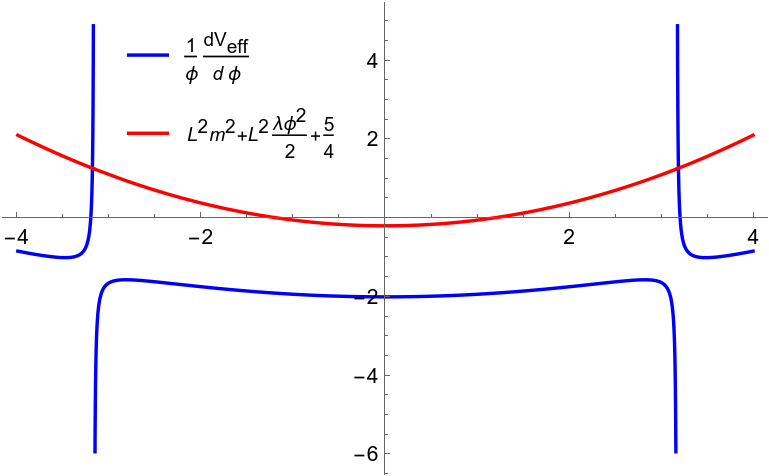} 
    \caption{} 
    \label{ads4b} 
    \vspace{1ex}
  \end{subfigure}
\end{minipage}
  \caption{AdS$_4$ single scalar: (a) Zero temperature $m^2-L$ phase plot for Neumann   boundary condition with $\lambda=0.4$ (b) Shows a point $m^2=-2, L=0.85, \l=0.4$ with the $\phi_{cl}\ne 0$ roots and the unitarity bound for Neumann boundary condition.} 
  \label{ads4} 
  \end{center} 
\end{figure}

\noindent
\textbf{Zero Temperature:}\\
The zero temperature phase plot is shown in Figure \ref{ads4ssmL}. Again as $M^2 L^2\rightarrow 0$, $\frac{1}{\phi_{cl}}\frac{\partial V(\phi_{cl})}{{\partial \phi_{cl}}} \rightarrow \infty$ because of the first $\tan$ term thus the equation (\ref{ads4sprime}) will not be satisfied. The boundary between regions $A-B$ is given by the BF bound $m^2 L^2=-9/4$  while between regions $B-C$ is given by the unitarity bound $m^2 L^2 =-5/4$. Even though region $B$ exhibits the symmetry preserving phase we get an unbounded potential as we cannot access $M^2=0$ due to the unitarity bound. Figure \ref{ads4b} shows a plot of $\frac{1}{\phi_{cl}}\frac{\partial V(\phi_{cl})}{{\partial \phi_{cl}}}$ and $M^2 L^2$ versus $\phi_{cl}$. The derivative of the potential has infinite number of disconnected branches with corresponding zeros coming from the $\tan$ function. Now the central curve also does not satisfy the unitarity bound.

\noindent
\textbf{Finite Temperature:}\\
The symmetry preserving phase observed at finite temperature is unstable as $M^2=0$ cannot be accessed due to the unitarity bound.

\subsubsection{AdS$_5$}

We shall now provide the expression for the effective potential for the theory on AdS$_5$. Setting $d=4$ in (\ref{trzt}) gives

\beqa\label{trzt5}
\f{1}{2 L^2}\mbox{tr}\left[\f{1}{-\square_E+V^{''}(\phi_{cl})}\right]=\f{{\cal V}_5}{L^524 \pi^2}\left(\f{\nu^5}{5} -\f{\nu^{3}}{3} \right)
\eeqa

where $\n$ and the effective mass, $M$ are related in this case as

\begin{equation}
\n=\sqrt{4+M^2L^2}=\left[4+\left(\f{\lambda}{2}\phi^{2}_{cl}+m^{2}\right)L^2\right]^{1/2}.
\end{equation}
The final expression for the one-loop corrected effective potential at zero temperature for Dirichlet and Neumann boundary conditions is given by

\begin{eqnarray}\label{ads5zero}
V^{(0)}_{eff}(\phi^{}_{cl})=\frac{1}{2} m^{2}\phi^{2}_{cl}+\frac{\lambda}{4!}\phi^{4}_{cl} \mp\frac{1}{24 L^5\pi^2}\left(\f{\n^5}{5}-\f{\n^3}{3} \right)
\end{eqnarray}

We now set the angular potential $\theta=0$ again and insert ${\cal V}_{5}=3\beta \pi^2 L^4/16$ at finite temperature to get the following effective potential
\begin{eqnarray}\label{ads5finitet}
V_{eff}(\phi^{}_{cl})=\frac{1}{2} m^{2}\phi^{2}_{cl}+\frac{\lambda}{4!}\phi^{4}_{cl}\mp\frac{1}{24 L^5\pi^2}\left(\f{\n^5}{5}-\f{\n^3}{3} \right)+\f{16}{3L^5\pi^2  \f{\b}{L}}\sum _{n=1}^{\infty} \frac{ e^{-\f{\b}{L} n (2\pm\n)}}{ n \left(1-e^{-\f{\b}{L} n}\right)^4}
\end{eqnarray}
The upper and the lower signs in the above expressions are for Dirichlet and Neumann boundary conditions respectively.
We further have in this case
\beqa
\D_\pm=2\pm\sqrt{4+M^2L^2}.
\eeqa
Thus the unitarity and BF bounds are  $M^2 L^2 \geq -3$ and $M^2 L^2 \geq -4$ respectively.

\subsection{$O(N)$ vector model}\label{onmodel}

The Lagrangian for the $O(N)$ vector model is given by

\beqa\label{onlag}
\mathcal{L}_E=\frac{1}{2} ( \partial ^{}_{\mu} \phi^i)^{2} + \frac{1}{2} m^{2} (\phi^i)^2 + \frac{\lambda}{4} \left[(\phi^i)^2\right ]^{2}~~~~~\mbox{where}~~~i=1,\cdots,N~.
\eeqa
We get a modified Klein-Gordon operator $[-\square^{}_{E}+M^{2}_{i}L^2]$ after expanding about the classical field as $\phi^i=\phi_{cl}^i+\eta^i$ and setting $\phi^{i}_{cl}=(0,0,\cdots,0,\phi^{}_{cl})$, with
\begin{equation}\label{effONmass}
M^{2}_{i}=
\begin{cases}
\lambda\phi^{2}_{cl}+m^{2}, & \text{for}\ \eta^{1}\cdots\eta^{N-1} \\
3\lambda\phi^{2}_{cl}+m^{2}, & \text{for}\ \eta^{N}
\end{cases} 
\end{equation}
The effective potential for this theory thus becomes
\begin{equation}
V^{}_{eff}(\phi^{}_{cl})=V(\phi^{}_{cl})+\frac{1}{2{\cal V}^{}_{d+1}}\bigg[{(N-1)}\mbox{tr} \log[-\square^{}_{E} L^2+M_1^{2}L^2]+\mbox{tr} \log[-\square^{}_{E} L^2+M_2^{2}L^2]\bigg]
\end{equation}

\subsubsection{AdS$_2$}
We give the expression for the effective potential renormalized including counterterms below
\beqa
V^{}_{eff}(\phi^{}_{cl})&=&\frac{1}{2} m^{2}\phi^{2}_{cl}+\frac{\lambda}{4!}\phi^{4}_{cl}
+\frac{(N-1)}{4 \pi L^2}\int^{M_1^{2}L^2}_{0} d x_1^2 \left[\psi^{(0)}\left(\sqrt{\f{1}{4}+m^2L^2}+\frac{1}{2}\right)\right.\nonumber \\
&-&\left.\psi^{(0)}\left(\sqrt{\f{1}{4}+x_1^2}+\frac{1}{2}\right)+\pi \tan\left(\pi \sqrt{\f{1}{4}+x_1^2}\right)-\pi \tan\left(\pi\sqrt{\f{1}{2}+m^2L^2} \right) \right]\nonumber\\
&+&\frac{1}{4 \pi L^2}\int^{M_2^{2}L^2}_{0} d x_2^2 \left[\psi^{(0)}\left(\sqrt{\f{1}{4}+m^2L^2}+\frac{1}{2}\right)-\psi^{(0)}\left(\sqrt{\f{1}{4}+x_2^2}+\frac{1}{2}\right)\right.\nonumber \\
&+&\left.\pi \tan\left(\pi \sqrt{\f{1}{4}+x_2^2}\right)-\pi \tan\left(\pi\sqrt{\f{1}{2}+m^2L^2} \right) \right]\nonumber\\
&+&\f{(N-1) }{ L^2\f{\b}{L}}\sum^{\infty}_{n=1}\frac{1}{n}\frac{e^{-n\f{\beta}{L}\left(\frac{1}{2}+\sqrt{\frac{1}{4}+M_1^2L^2}\right)}}{\left|1-e^{-n\f{\beta}{L}}\right|}+\f{1 }{ L^2\f{\b}{L}}\sum^{\infty}_{n=1}\frac{1}{n}\frac{e^{-n\f{\beta}{L}\left(\frac{1}{2}+\sqrt{\frac{1}{4}+M_2^2L^2}\right)}}{\left|1-e^{-n\f{\beta}{L}}\right|}
\eeqa
where $M_i$ are defined in \ref{effONmass}. The phase plots in this scheme are once again similar to single scalar.

We next study the variation of phases with $N$ by writing the effective potential renormalized using the $\overline{\text{MS}}$ scheme and  introducing the scale $\mu$ at zero temperature. In this scheme we have removed the finite $\g + \log(4\pi)$ along with the $1/\epsilon$ piece from the trace in \ref{expads2} to get

\beqa
V^{}_{eff}(\phi^{}_{cl})&=&\frac{1}{2} m^{2}\phi^{2}_{cl}+\frac{\lambda}{4!}\phi^{4}_{cl} \\
&+&\frac{(N-1)}{4 \pi L^2}\int^{M_1^{2}L^2}_{0} d x_1^2 \left[\log(\mu^2 L^2)-\psi^{(0)}\left(\sqrt{\f{1}{4}+x_1^2}+\frac{1}{2}\right)+\pi \tan\left(\pi \sqrt{\f{1}{4}+x_1^2}\right) \right]\nonumber\\
&+&\frac{1}{4 \pi L^2}\int^{M_2^{2}L^2}_{0} d x_2^2 \left[\log(\mu^2 L^2)-\psi^{(0)}\left(\sqrt{\f{1}{4}+x_2^2}+\frac{1}{2}\right)+\pi \tan\left(\pi \sqrt{\f{1}{4}+x_2^2}\right)\right] \nonumber
\eeqa

\begin{figure}[H] 
\begin{center} 
  \begin{minipage}{0.45\textwidth}%
   \begin{subfigure}[b]{0.9\linewidth}
    \centering
    \includegraphics[width=1\linewidth]{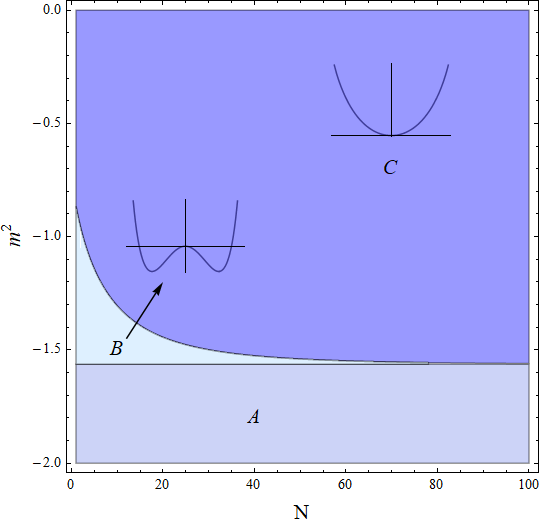} 
    \caption{} 
    \label{mvsnd} 
    \vspace{1ex}
  \end{subfigure}
  \end{minipage}
 \begin{minipage}{0.45\textwidth}%
   \begin{subfigure}[b]{0.9\linewidth}
    \centering
    \includegraphics[width=1\linewidth]{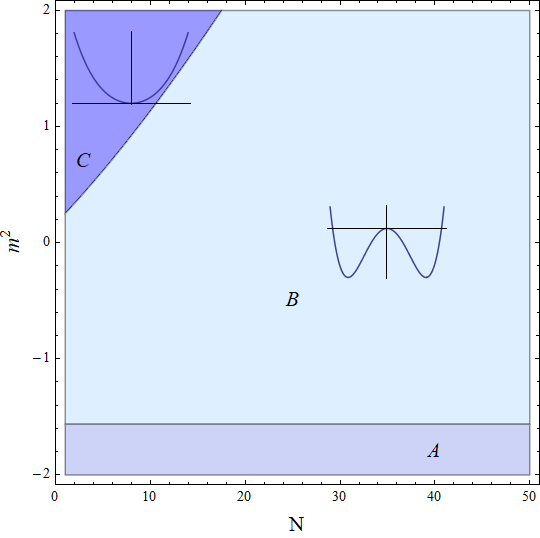} 
    \caption{} 
    \label{mvsnn} 
    \vspace{1ex}
  \end{subfigure}
\end{minipage}
  \caption{AdS$_2$ $O(N)$ model: Phase plots with $\lambda=0.4, L=0.4$ at zero temperature for (a) Neumann boundary condition and (b) Dirichlet boundary condition } 
  \label{mvsN} 
  \end{center} 
\end{figure}

We observe that the symmetry breaking phase shrinks as $N$ increases and ceases to exist beyond a particular value of $N$ for the Neumann case while for the Dirichlet case the symmetry breaking phase persists even as we increase $N$ as shown in Figures \ref{mvsnd} and \ref{mvsnn} respectively. The region $A$ in these plots falls below the BF bound while the boundary between $B-C$ represents the second order transition. The characteristics of the phase diagram shown in Figure \ref{mvsnd} differ from that of the phase diagram in Figure \ref{ads2ssmL} depending on the value of $\mu$ selected.

\subsubsection{AdS$^{}_{3}$}\label{ads3fn}

Similar to the case of the single scalar theory, where the trace is given by (\ref{trzt3}), the leading contribution to the effective potential for the $O(N)$ vector model including the expression for the partition function (\ref{finalp3n}) can be written as

\begin{eqnarray}
V^{}_{eff}(\phi^{}_{cl})&=&\frac{1}{2}m^{2}\phi^{2}_{cl}+\frac{\lambda}{4}\phi_{cl}^{4}\mp
\frac{1}{12\pi L^3}\big[(N-1)\n(M_1^{2}L^2)^{3} +\n(M_2^{2}L^2)^{3}\big] \non
&-&\f{(N-1)}{{\cal V}_3}\sum^{\infty}_{n=1}\frac{1}{n}\frac{e^{-n\f{\beta}{L}\left(1\pm\n(M_1^2 L^2)\right)}}{|1-e^{-n\f{\b}{L}}|^{2}}-\f{1}{{\cal V}_3}\sum^{\infty}_{n=1}\frac{1}{n}\frac{e^{-n\f{\beta}{L}\left(1\pm\n(M_2^2 L^2)\right)}}{|1-e^{-n\f{\b}{L}}|^{2}}
\end{eqnarray}

where $\n(M_i^2 L^2)=\sqrt{1+M_i^2 L^2}$ and the upper and the lower signs in the above expressions are again for Dirichlet and Neumann boundary conditions respectively. Extremizing the potential gives,

\begin{eqnarray}
0=\frac{\partial V}{\partial \phi^{}_{cl}}=\l \phi^{3}_{cl}+m^{2}\phi^{}_{cl}\mp\frac{ (N-1)\l \phi^{}_{cl} \n(M_1^{2}L^2)}{4 \pi L}\mp\frac{3\l\phi^{}_{cl} \n(M_2^2 L^2)}{4 \pi L}\non
-\sum ^{\infty}_{n=1} \frac{ (N-1)\b L\l\phi^{}_{cl} e^{- n \f{\b}{L}  \left(1\pm\n(M_1^{2}L^2)\right)}}{{\cal V}_3    \left(1-e^{-n\f{\b}{L} }\right)^2 \n(M_1^{2}L^2)}
-\sum ^{\infty}_{n=1} \frac{3\b L \l \phi^{}_{cl} e^{- n \f{\b}{L}\left(1\pm\n(M_2^{2}L^2)\right)}}{{\cal V}_3   \left(1-e^{- n \f{\b}{L}}\right)^2 \n(M_2^{2}L^2)}
\end{eqnarray}

\begin{figure}[H] 
\begin{center} 
  \begin{minipage}{0.45\textwidth}%
   \begin{subfigure}[b]{0.9\linewidth}
    \centering
    \includegraphics[width=1\linewidth]{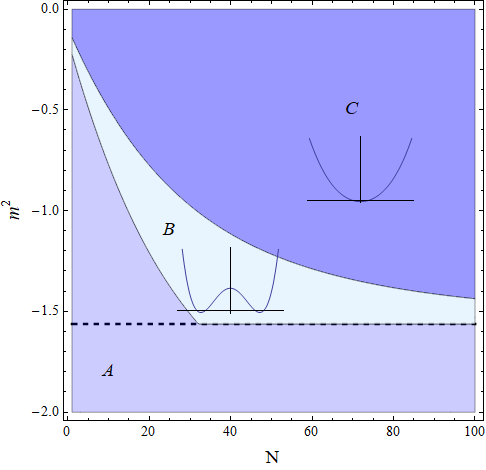} 
    \caption{} 
    \label{ads3onn} 
    \vspace{1ex}
  \end{subfigure}
  \end{minipage}
 \begin{minipage}{0.45\textwidth}%
   \begin{subfigure}[b]{0.9\linewidth}
    \centering
    \includegraphics[width=1\linewidth]{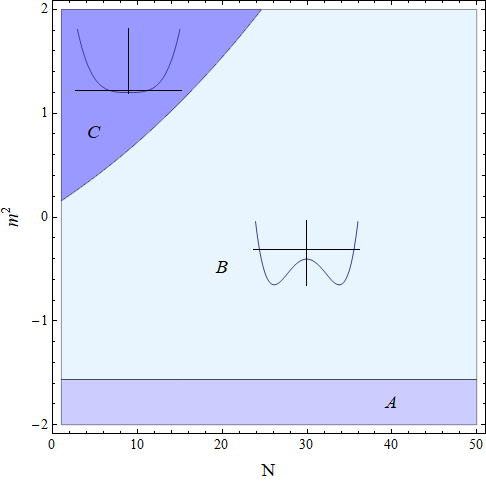} 
    \caption{} 
    \label{ads3ond} 
    \vspace{1ex}
  \end{subfigure}
\end{minipage}
  \caption{AdS$_3$ $O(N)$ model: Phase plots for $m^2$ vs $N$ with $\l=0.5, L=0.8$ for (a) Neumann boundary condition (b) Dirichlet boundary condition.} 
  \label{ads3on} 
  \end{center} 
\end{figure}

The $m^2-L$ phase plots are qualitatively similar to single scalar. Further Figures \ref{ads3onn} and \ref{ads3ond} shows the $m^2 - N$ plots for Neumann and Dirichlet boundary cases respectively. Region $A$ is disallowed as it is below the BF bound in \ref{ads3ond} while in Figure \ref{ads3onn} it is the combined region where both unitarity and the BF bound are violated with the dashed black line demarcating the region below the BF bound. Region $B$ does not vanish for the Neumann boundary condition case as $N$ is increased in contrast to the case in AdS$_2$.

\subsubsection{AdS$_4$}
Expanding the expression for the trace at zero temperature around $d=3$ gives
\begin{eqnarray}\label{zads4exp}
\frac{1}{{\cal V}^{}_{d+1} L^2}\mbox{tr} \frac{1}{-\square+M_i^{2}}&=&\frac{(2+M_i^2L^2)}{16\pi^2 L^4}\left[-\frac{2}{\e}-1+\gamma-\log(4\pi)+\psi^{(0)}\left(\n(M_i^2 L^2)-\f{1}{2}\right)\right.\nonumber\\
&+&\psi^{(0)}\left(\n(M_i^2 L^2)+\f{3}{2}\right)\Bigg]
\end{eqnarray}

where $\n(M_i^2 L^2)=\sqrt{9/4+M_i^2 L^2}$ and $\e=3-d$. The effective potential is renormalized by adding counterterms corresponding to $m^2$ and $\l$ 
so that,

\begin{eqnarray}\label{effnct4}
V_{eff}(\phi_{cl}) =\frac{1}{2}m^{2}\phi^{2}_{cl}+\frac{\lambda}{4}\phi_{cl}^{4}-\f{1}{{\cal V}_{d+1}}(\log Z^{(1)}+\log Z^{(1)}_{\b})
+ \f{\d m^2}{2} \phi_{cl}^2+\f{\d\l}{4}\phi_{cl}^4
\end{eqnarray}

where the zero temperature contribution is,

\beqa
-\f{1}{{\cal V}_{d+1} L^2}\log Z^{(1)}=\f{\l}{{2\cal V}_{d+1} L^2}\int_0^{\phi_{cl}^2}\left[(N-1)\mbox{tr} \frac{1}{-\square+M^{2}_1}+3\mbox{tr} \frac{1}{-\square+M^{2}_2}\right]d\phi_{cl}^2 .
\eeqa

To renormalize we set the following renormalization conditions (at zero temperature),

\beqa\label{renorcads4}
\left.\frac{\partial }{\partial \phi_{cl}^{2}}V^{0}_{eff}(\phi_{cl})\right|_{\phi_{cl}=0}=\f{m^2}{2}
~~~~~~~
\left.\frac{\partial^2 }{\partial (\phi_{cl}^2)^2}V^{0}_{eff}(\phi_{cl})\right|_{\phi_{cl}=0}=\f{\l}{2}
\eeqa

which give

\beqa
\d m^2=-\l(N+2)\f{1}{{\cal V}_{d+1}}\mbox{tr} \frac{1}{-\square+m^{2}}~\mbox{and}~\d \l=-\l^2(N+8)\f{1}{{\cal V}_{d+1}}\f{\partial}{\partial m^2}\mbox{tr} \frac{1}{-\square+m^{2}}
\eeqa
The renormalized effective potential at zero temperature, after removing (an infinite) constant thus becomes

\begin{eqnarray}\label{effnct4ren}
V_{eff}(\phi_{cl}) &=&\frac{1}{2}m^{2}\phi^{2}_{cl}+\frac{\lambda}{4}\phi_{cl}^{4}+ \f{\l}{2\mathcal{V}_4}\int_0^{\phi_{cl}^2}\left[(N-1)\mbox{tr} \frac{1}{-\square+M^{2}_1}+3\mbox{tr} \frac{1}{-\square+M^{2}_2 }\right]_{\mbox{ren}}d\phi_{cl}^2 \nonumber \\
&-&\f{\l^2(N+8)}{128\pi^2}\f{(2+m^2 L^2)}{\n(m^2 L^2)}\left[\psi^{(1)}\left(\n(m^2 L^2)-\f{1}{2}\right)+\psi^{(1)}\left(\n(m^2 L^2)+\f{3}{2}\right)\right.\nonumber\\
&-&2\pi^2 \sec^2\left(\pi \nu(m^2 L^2)\right)\Bigg]\phi_{cl}^4 
\end{eqnarray}

where

\beqa
&&\f{1}{{\cal V}_{4}L^2}\left[(N-1)\mbox{tr} \frac{1}{-\square+M^{2}_1 }+3\mbox{tr} \frac{1}{-\square+M^{2}_2}\right]_{\mbox{ren}}\non
&=&
(N-1)\frac{(2+M_1^2 L^2)}{16\pi^2 L^4}\left[\psi^{(0)}\left(\n(M_1^2 L^2)-\f{1}{2}\right)+\psi^{(0)}\left(\n(M_1^2 L^2)+\f{3}{2}\right)-\psi^{(0)}\left(\n(m^2 L^2)-\f{1}{2}\right)\right.\nonumber\\
&-&\left.\psi^{(0)}\left(\n(m^2 L^2)+\f{3}{2}\right)-2\pi \tan(\pi \nu(M_1^2 L^2))+2\pi \tan(\pi \nu(m^2 L^2))\right]\nonumber\\
&+&3\frac{(2+M_2^2 L^2)}{16\pi^2 L^4}\left[\psi^{(0)}\left(\n(M_2^2 L^2)-\f{1}{2}\right)+\psi^{(0)}\left(\n(M_2^2 L^2)+\f{3}{2}\right)-\psi^{(0)}\left(\n(m^2 L^2)-\f{1}{2}\right)\right.\nonumber\\
&-&\left.\psi^{(0)}\left(\n(m^2 L^2)+\f{3}{2}\right)-2\pi \tan\left(\pi \nu(M_2^2 L^2)\right)+2\pi \tan\left(\pi \nu(m^2 L^2)\right)\right] \nonumber\\
&-&\sum _{n=1}^{\infty} \frac{3 (N-1) e^{-n\f{\b}{L}\left(\f{3}{2}-\nu(M_1^2 L^2) \right)}}{2 L^4 (\b/L) \pi \left|1-e^{-n\f{\b}{L}} \right|^3}-\sum _{n=1}^{\infty} \frac{9 e^{-n\f{\b}{L}\left(\f{3}{2}-\nu(M_2^2 L^2) \right)}}{2 L^4 (\b/L) \pi \left|1-e^{-n\f{\b}{L}} \right|^3}
\eeqa
We observe here that the symmetry breaking region does not exist and we get an unstable symmetry preserving phase like in the single scalar case.

\subsubsection{AdS$_5$}
We also give the final expression for AdS$_5$ by expanding the expression for the trace in \ref{trzt} at zero temperature around $d=4$ below
\begin{eqnarray}
V^{}_{eff}(\phi^{}_{cl})&=&\frac{1}{2}m^{2}\phi^{2}_{cl}+\frac{\lambda}{4}\phi_{cl}^{4}\\
&\mp&
\frac{1}{12\pi L^5}\left[(N-1)\left(\f{\n^5(M_1^2L^2)}{5}-\f{\n^3(M_1^2L^2)}{3} \right) +3\left(\f{\n^5(M_2^2L^2)}{5}-\f{\n^3(M_2^2L^2)}{3} \right)\right] \non
&+&\f{(N-1)}{{\cal V}_5}\sum^{\infty}_{n=1}\frac{1}{n}\frac{e^{-n\f{\beta}{L}\left(2\pm \n(M_1^2L^2)\right)}}{\left|1-e^{-n\f{\b}{L}}\right|^{4}}+\f{3}{{\cal V}_5}\sum^{\infty}_{n=1}\frac{1}{n}\frac{e^{-n\f{\beta}{L}\left(2\pm \n(M_2^2 L^2)\right)}}{\left|1-e^{-n\f{\b}{L}}\right|^{4}}\nonumber
\end{eqnarray}
where $\n(M_i^2 L^2)=\sqrt{4+M_i^2 L^2}$ and again the upper and the lower signs in the above expression are for Dirichlet and Neumann boundary conditions respectively.

\subsection{Large $N$}
We now present some basics using the notation as in \cite{Carmi:2018qzm}. The Lagrangian for the $O(N)$ vector model is given by
(\ref{onlag}). We re-scale the coupling $\l\rightarrow \l/N$ and introduces an auxiliary field $\s$ for organizing the perturbation theory in $1/N$ so that the Lagrangian is

\begin{equation}
\mathcal{L}=\frac{1}{2}(\partial_{\m}\phi^{i})^{2}+\frac{m^{2}}{2}(\phi^{i})^{2}-\frac{1}{2\lambda}\sigma^{2}+\frac{1}{\sqrt{N}}\sigma(\phi^{i})^{2}.
\end{equation}

We again expand the fields as $\phi^i(x)=\sqrt{N}\phi_{cl}^i+\d\phi^i(x)$ and $\s(x)=\sqrt{N}\s_{cl}+\d\s(x)$, and then perform the integral over the fluctuations $\d\phi^i(x)$ to get the following effective potential to the leading order in $1/N$

\begin{equation}
V_{eff}(\phi_{cl}^i,\s_{cl})=N\left[\frac{M^{2}}{2}(\phi^{i}_{cl})^{2}-\frac{(M^{2}-m^2)^2}{8\l}+\frac{1}{2 \mathcal{V}_{d+1}} \mbox{tr} \log \left(-\square^{}_{E} L^2 +M^2 L^2\right)\right].
\end{equation}
where $M^2=m^2+2\s_{cl}$. Separating the contributions from the zero and finite temperature we have the following effective potential

\begin{equation}\label{effln}
V_{eff}(\phi_{cl}^i,\s_{cl})=N\left[\frac{M^{2}}{2}(\phi^{i}_{cl})^{2}-\frac{(M^{2}-m^2)^2}{8\l}-\f{1}{{\cal V}_{d+1}}(\log Z^{(1)}+\log Z^{(1)}_{\t})\right].
\end{equation}

In the following subsections we shall study the phases in various dimensions.

\subsubsection{AdS$_2$} \label{ads2largeNsection}
The zero temperature contribution to the trace can be obtained from (\ref{trzt}). The trace being divergent we expand about $d=1$ as in (\ref{expads2}) and to renormalize the effective potential we include a mass counterterm so that

\begin{equation}\label{efflnct}
\f{V_{eff}(\phi_{cl}^i,\s_{cl})}{N}=-\frac{(M^{2}-m^2)^2}{8\l}+\frac{M^{2}}{2}(\phi^{i}_{cl})^{2}+M^2L^2\d m^2-\f{1}{{\cal V}_{d+1}}(\log Z^{(1)}+\log Z^{(1)}_{\t}).
\end{equation}
and impose 

\beqa
\left.\frac{1}{N}\frac{\partial }{\partial M^{2}}V^{0}_{eff}(\phi_{cl}^i,\s_{cl})\right|_{\sigma=\phi_{cl}^i=0}=0.
\eeqa
where $V^{0}_{eff}$ in this equation is for zero temperature. This renormalization condition gives

\beqa
\d m^2&=&-\f{1}{2{\cal V}_{d+1}L^2}\tr \left[\frac{1}{-\square_E+m^2}\right]\\
&=&-\f{1}{4\pi\e L^2}-\f{1}{8\pi L^2}\left[-2\psi^{(0)}\left(\f{1}{2}+\sqrt{\f{1}{4}+m^2 L^2}\right) +\log (4\pi) -\g+2\pi\tan\left(\pi \sqrt{\f{1}{4}+m^2 L^2}\right)\right] \nonumber
\eeqa
Inserting the finite temperature contribution from (\ref{zdto1}), ${\cal V}^{}_{2}=-\beta L$ the effective potential at finite temperature becomes
\begin{eqnarray}
\frac{V^{}_{eff}}{N}&=&-\frac{(M^{2}-m^{2})^{2}}{8\lambda}+\frac{1}{2}M^{2}(\phi^{i}_{cl})^{2}-\frac{1}{4\pi L^2}\int^{M^{2} L^2}_{0} d(M^2 L^2) \left[ \psi^{(0)}\left(\n+ \frac{1}{2}\right)-\pi\tan(\pi \nu)\right.\nonumber\\
&-&\psi^{(0)}\left(\f{1}{2}+\sqrt{\f{1}{4}+m^2 L^2}\right)+\left.\pi\tan\left(\pi \sqrt{\f{1}{4}+m^2 L^2}\right)\right]+\f{1}{\f{\b}{L} L^2}\sum^{\infty}_{n=1}\frac{1}{n}\frac{e^{-n\f{\beta}{L}\left(\frac{1}{2}-\sqrt{\frac{1}{4}+M^2 L^2}\right)}}{\left|1-e^{-n\f{\beta}{L}}\right|} \nonumber \\
\end{eqnarray}
and the saddle point equations are

\begin{eqnarray}\label{sads2}
0&=&\frac{1}{N}\frac{\partial V_{eff}}{\partial M^{2}}=\frac{m^{2}-M^{2}}{4\lambda}+\f{(\phi^{i}_{cl})^{2}}{2}-\frac{1}{4\pi}\psi^{(0)} \left(\frac{1}{2}+\sqrt{\frac{1}{4}+M^{2}L^2}\right)+\f{1}{4}\tan\left(\pi \sqrt{\frac{1}{4}+M^{2}L^2}\right)\nonumber\\
&+&\f{1}{4\pi}\psi^{(0)}\left(\f{1}{2}+\sqrt{\f{1}{4}+m^2 L^2}\right)-\f{1}{4}\tan\left(\pi \sqrt{\f{1}{4}+m^2 L^2}\right)\nonumber\\
&+&\f{1}{2}\sum^{\infty}_{n=1}\frac{e^{-n\f{\beta}{L}\left(\frac{1}{2}-\sqrt{\frac{1}{4}+M^2 L^2}\right)}}{\left|1-e^{-n\f{\beta}{L}}\right|\sqrt{\frac{1}{4}+M^{2}L^2}} \\
0&=&\frac{1}{N}\frac{\partial V}{\partial (\phi_{cl}^i)^{2}}=\phi_{cl}^i M^2
\end{eqnarray}\\

\noindent
\textbf{Zero Temperature:}\\
We will have no symmetry breaking solutions when $M^2=0$ at zero temperature itself as when $M^2 L^2\rightarrow 0$, $\frac{1}{N}\frac{\partial V_{eff}}{{\partial M^2}} \rightarrow \infty$ because of the first $\tan$ term and thus the saddle point equation has no solution. Further even though the symmetry preserving phase exists, the potential in this case is unbounded below as $\frac{1}{N}\frac{\partial^2 V}{\partial (\phi_{cl}^i)^2}=M^2<0$ because of the unitarity bound.\\

\noindent
\textbf{Finite Temperature:}\\
Further extending to finite temperature we observe that the symmetry preserving phase is unstable because of the requirement of convergence of the finite temperature contribution to the effective potential requiring $\b(d/2-\nu)>0$ or $M^2<0$ and preventing us from accessing $M^2=0$. This places a more stringent constraint than just unitarity.

\subsubsection{AdS$_3$}\label{ads3n}

As in the case of the single scalar theory, with the trace given by (\ref{trzt3}), the leading contribution to the effective potential for Dirichlet/Neumann boundary conditions for the large $N$ theory (\ref{effln}) at zero temperature, after removing an infinite constant is,
 
\begin{equation}
\label{e122}
\frac{V^0_{eff}(M^{2},\phi^{i}_{cl})}{N}=-\frac{\left(M^{2}-m^{2}\right)^{2}}{8\lambda}+\frac{1}{2}(\phi^{i}_{cl})^{2}M^{2}\mp\frac{(1+M^{2}L^2)^{\frac{3}{2}}}{12\pi L^3}
\end{equation}
where the upper sign is for Dirichlet boundary condition and lower sign for Neumann boundary condition.
The saddle point equations are
\begin{eqnarray}\label{saddle3}
0=\frac{1}{N}\frac{\partial V_{eff}}{\partial M^{2}}=\frac{m^{2}-M^{2}}{4\lambda}+\frac{(\phi^{i}_{cl})^{2}}{2}\mp\frac{\sqrt{1+M^{2} L^2}}{8\pi L}
\end{eqnarray}

\beqa
0=\frac{1}{N}\frac{\partial V_{eff}}{\partial (\phi_{cl}^i)^2}= \phi_{cl}^iM^2
\eeqa
Including the expression for the partition function (\ref{finalp3n}) at finite temperature, 

\begin{eqnarray}
\frac{V^{}_{eff}(M^{2},\phi^{i}_{cl})}{N}=-\frac{\left(M^{2}-m^{2}\right)^{2}}{8\lambda}+\frac{1}{2}(\phi^{i}_{cl})^{2}M^{2}\mp\frac{(1+M^{2}L^2)^{\frac{3}{2}}}{12\pi L^3} -\f{1}{{\cal V}_3}\sum^{\infty}_{n=1}\frac{1}{n}\frac{e^{-n\f{\beta}{L}\left(1-\sqrt{1+M^2 L^2}\right)}}{\left|1-e^{-n\f{\b}{L}}\right|^{2}} \nonumber \\
\end{eqnarray}

The first saddle point equation is
\begin{eqnarray}\label{saddle3t}
0=\frac{1}{N}\frac{\partial V_{eff}}{\partial M^{2}}=\frac{m^{2}-M^{2}}{4\lambda}+\frac{(\phi^{i}_{cl})^{2}}{2}-\frac{\sqrt{1+M^{2} L^2}}{8\pi L}-\f{1}{{\cal V}_3}\sum^{\infty}_{n=1}\frac{\b L e^{-n\f{\beta}{L}\left(1-\sqrt{1+M^2 L^2}\right)}}{2\left|1-e^{-n\f{\b}{L}}\right|^{2}\sqrt{1+M^{2}L^2}}
\end{eqnarray}

\begin{figure}[H] 
\begin{center} 
  \begin{minipage}{0.45\textwidth}%
   \begin{subfigure}[b]{0.9\linewidth}
    \centering
    \includegraphics[width=1\linewidth]{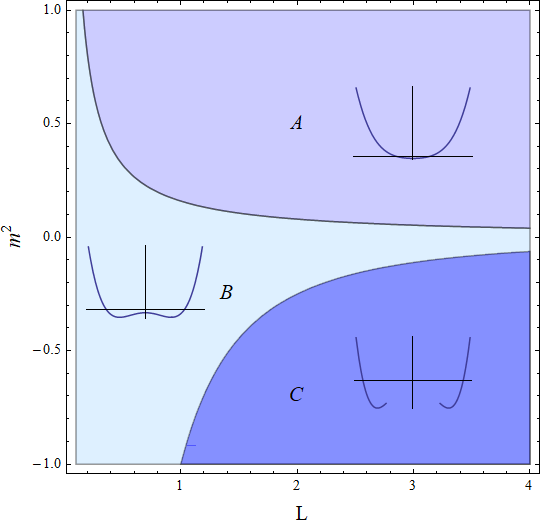} 
    \caption{} 
    \label{ads3lnd0} 
    \vspace{1ex}
  \end{subfigure}
  \end{minipage}
 \begin{minipage}{0.45\textwidth}%
   \begin{subfigure}[b]{0.9\linewidth}
    \centering
    \includegraphics[width=1\linewidth]{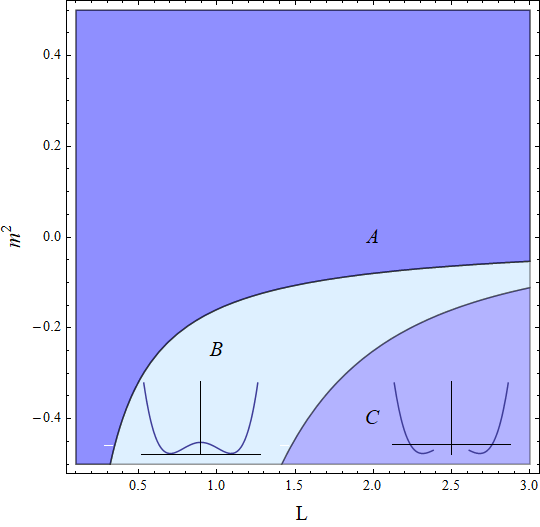} 
    \caption{} 
    \label{ads3lnn0} 
    \vspace{1ex}
  \end{subfigure}
\end{minipage}
  \caption{AdS$_3$ large $N$: Zero temperature $m^2-L$ phase plot for (a) Dirichlet and (b) Neumann boundary conditions with $\lambda=1$} 
  \label{largeN0} 
  \end{center} 
\end{figure}

\noindent
\textbf{Zero Temperature:}\\
\noindent
To check for the existence of the symmetry breaking phase we put $M^2=0$ and get for Dirichlet /Neumann the range
\begin{eqnarray}\label{ads3lnbound}
    -\f{1}{L^2}\leq m^2 \leq\pm\f{\l}{2 \pi L}
\end{eqnarray}
respectively. The $m^2-L$ phase plots for Dirichlet and Neumann boundary conditions are shown in Figures \ref{ads3lnd0} and \ref{ads3lnn0} respectively. The equations (\ref{ads3lnbound}) are also the boundaries between regions $B-C$ and $A-B$ respectively in the $m^2 - L$ phase plots shown in Figure \ref{largeN0}. Region $A$ (which would have otherwise included the symmetry preserving phase) in Figure \ref{ads3lnn0} falls outside the unitarity bound while the potenatial plots get cut at smaller values of $\phi_{cl}$ in region $C$ of both the plots.

\noindent
\textbf{Finite Temperature:}\\
We  notice that the above finite temperature series converges only when in the last temperature dependent term we have
\beqa \label{convergence}
\b(d/2-\nu) >0 
\eeqa
and thus requires $\nu<d/2$ or $M^2<0$. So series convergence places a stricter constraint than unitarity. Even though in this case the effective potential reduces to the zero temperature effective potential as $\b \rightarrow \infty$, the symmetry breaking solution of $M^2=0$ cannot be accessed due to the constraint in equation (\ref{convergence}).

\subsubsection{AdS$_4$}\label{ads4largeNsection}
We may write the effective potential as
\begin{eqnarray}\label{efflnct4}
\f{V_{eff}(\phi_{cl}^i,\s_{cl})}{N} &=&-\frac{(M^{2}-m^2)^2}{8\l}+\frac{M^{2}}{2}(\phi^{i}_{cl})^{2}-\f{1}{{\cal V}_{d+1}}(\log Z^{(1)}+\log Z^{(1)}_{\b})\non
&+& M^2\d m^2-\f{(M^2-m^2)^2}{2}\d\left(\f{1}{\l}\right).
\end{eqnarray}

With the renormalization conditions (at zero temperature),

\beqa
\left.\frac{1}{N}\frac{\partial }{\partial M^{2}}V^{0}_{eff}(\phi_{cl}^i,\s_{cl})\right|_{\s_{cl}=\phi_{cl}^i=0}=0
\eeqa
\beqa
\left.\frac{1}{N}\frac{\partial^2 }{\partial (M^{2})^2}V^{0}_{eff}(\phi_{cl}^i,\s_{cl})\right|_{\s_{cl}=\phi_{cl}^i=0}=-\f{1}{16\l}
\eeqa

we can compute the counterterms to be

\beqa
\d m^2&=&-\f{1}{2{\cal V}_{d+1}}\tr \left[\frac{1}{-\square_E+M^2}\right]_{\s_{cl}=\phi_{cl}^i=0}
\eeqa

\beqa
\d\left(\f{1}{4\l}\right)&=&\f{1}{2{\cal V}_{d+1}}\f{\pa}{\pa M^2}\tr \left[\frac{1}{-\square_E+M^2}\right]_{\s_{cl}^2=\phi_{cl}^i=0}
\eeqa

We can now write down the renormalized effective potential at finite temperature,

\beqa\label{vads4ln}
\frac{V(M^{2},\phi^{i}_{cl})}{N}&=&-\frac{(M^{2}-m^{2})^{2}}{8\lambda}+\frac{1}{2}M^{2}(\phi^{i}_{cl})^{2}+\int_0^{M^2 L^2}\f{(2+M^2 L^2)}{32 \pi^2 L^4}\left[ \psi^{(0)}\left(\f{3}{2}+\nu(M^2L^2)\right)\right.\nonumber\\
&+&\psi^{(0)}\left(-\f{1}{2}+\nu(M^2L^2)\right)-\psi^{(0)}\left(\f{3}{2}+\nu(m^2L^2)\right)-\psi^{(0)}\left(-\f{1}{2}+\nu(m^2L^2)\right)\nonumber\\
&-&2\pi \tan\left(\pi \nu(M^2L^2)\right)+2\pi\tan\left(\pi \nu(m^2L^2)\right) \Bigg] d(M^2 L^2)\nonumber\\
&-&\f{(2+m^2 L^2)}{128 \pi^2 ~ \nu(m^2L^2)}\left(  \psi^{(1)}\left(\f{3}{2}+\nu(m^2L^2)\right)+\psi^{(1)}\left(-\f{1}{2}+\nu(m^2L^2)\right)\right.\nonumber\\
&-&2\pi^2\sec^2\left(\pi \nu (m^2 L^2)\right)\Bigg)(M^2-m^2)^2 -\sum _{n=1}^{\infty} \frac{3 e^{-n\f{\b}{L}\left(\f{3}{2}-\nu(M^2 L^2) \right)}}{2 L^4 \f{\b}{L} \pi \left|1-e^{-n\f{\b}{L}} \right|^3}
\eeqa

\noindent
\textbf{Zero Temperature:}\\
Again the symmetry breaking solution of $M^2=0$ is not possible because of the constrain put by the unitarity bound and the symmetry preserving phase will be unstable  as $\frac{1}{N}\frac{\partial^2 V}{\partial (\phi_{cl}^i)^2}=M^2<0$. \\

\noindent
\textbf{Finite Temperature:}\\
The finite temperature symmetry preserving phase too will be unstable because of the requirement of convergence of the finite temperature series that implies $M^2=0$ is not accessible.

\subsubsection{AdS$_5$}
 We also below provide the expression for the effective potential at finite temperature in AdS$_5$
\begin{equation}
\label{ads5ln}
\frac{V^0_{eff}(M^{2},\phi^{i}_{cl})}{N}=-\frac{\left(M^{2}-m^{2}\right)^{2}}{8\lambda}+\frac{1}{2}(\phi^{i}_{cl})^{2}M^{2}\mp \frac{1}{24 L^5\pi^2}\left(\f{\n^5}{5}-\f{\n^3}{3} \right)+\f{16}{3L^5\pi^2  \f{\b}{L}}\sum _{n=1}^{\infty} \frac{ e^{-\f{\b}{L} n (2\pm\n)}}{ n \left|1-e^{-\f{\b}{L} n}\right|^4}
\end{equation}
where now
$\nu=\sqrt{4+M^2 L^2}$ and the upper and the lower signs in the above expression correspond to Dirichlet and Neumann boundary conditions respectively.

\subsection{Behaviour of Correlators}
 To check consistency with our numerical results in the previous sections, we now in this section verify the existence of symmetry breaking phase by looking at two point correlators at zero temperature. In case the correlator is found to have a long range order then it supports second order phase transition indicating a symmetry breaking phase otherwise this phase is prohibited from existing for the given theory and dimension. We have the following relation between the trace and the two point function \cite{Inami:1985dj} \cite{Kakkar:2022hub}
 
\beqa
\tr\left[\f{1}{-\square_E+V''(\phi_{cl}) }\right]=\int d^{d+1} x \sqrt{g} ~ G(x,x)
\eeqa
we can thus write the effective potential for the large $N$ model as

\begin{equation}
V_{eff}(\phi_{cl}^i,\s_{cl})=N\left[\frac{M^{2}}{2}(\phi^{i}_{cl})^{2}-\frac{(M^{2}-m^2)^2}{8\l}+\frac{1}{2} \int_0^{M^2L^2}G\left(x,x;M^2L^2\right)d (M^2 L^2)\right].
\end{equation}
The scalar bulk to bulk propagator is given by \cite{Burgess:1984ti}, \cite{Inami:1985wu}

\beqa
G_\D(x_1,x_2)=\f{\G(\D)}{2 \pi^{\f{d}{2}}\G(\D-\f{d}{2}+1)}\f{L^{2\D-d+1}}{\xi(x_1,x_2)^\D}{}_2 F_1\left(\D,\D-\f{d}{2}+\f{1}{2},2\D-d+1,-\f{4 L^2}{\xi(x_1,x_2)}\right) \nonumber\\
\eeqa
where $\xi$ is  the squared chordal distance between the two points given in Poincar\'e coordinates as

\beqa
\xi(x_1,x_2)=L^2\f{(z_1-z_2)^2+(\vec{x}_1-\vec{x}_2)^2}{z_1 z_2}.
\eeqa
The behaviour of correlators has been used for studying symmetry breaking in the large $N$ model in AdS$_2$  \cite{Inami:1985dj} which is otherwise forbidden in flat space \cite{Mermin:1966fe}, \cite{Coleman:1973ci}.
Following \cite{Inami:1985dj}, for our analysis we consider the simple case of breakdown of $U(1)$ symmetry and look at the large distance behavior of $\langle\phi^*(x)\phi(0)  \rangle$ correlator. We parameterize $\phi(x)=\rho(x)e^{i\theta(x)}$ as invariant under $\th(x)\rightarrow \th(x)+\alpha$. The two point function is then given by

\beqa \label{correlator}
\langle\phi^*(x)\phi(x_0)  \rangle\simeq\rho_0^2 \exp\left[\langle\th(x)\th(x_0)  \rangle-\f{1}{2}\left(\langle\th^2(x)  \rangle+\langle\th^2(x_0)  \rangle\right)\right]
\eeqa
where $\rho_0$ is the expectation value of $\rho(x)$ for the corresponding theory. The coincident limit of the propagators can be read from the expressions of the trace we have computed in the previous sections. We now look at the large distance behavior of the propagators in the following sub-section.
\subsubsection{Large distance behavior of Propagator}
In the large $\xi$ limit the propagator behaves as

\beqa
G_\D(x_1,x_2)\sim \f{L^{2\D-d+1}\G(\D)}{2 \pi^{d/2}\G(\D-\f{d}{2}+1)} \f{1}{\xi(x_1,x_2)^\D} \sim  \f{C'}{\xi(x_1,x_2)^\D}
\eeqa

where $C'$ is the collection of non $\xi$ dependent terms including factors of $L$. We further introduce another parameter $C=C' L^{d-1}$ which will help us in writing the expressions for the correlators later.
Next we analyse the behavior of the propagator in different cases.\\

\noindent
(1) \textbf{Dirichlet Boundary Condition:}\\

\noindent
(a) \underline{Single Scalar}

\beqa \label{ld1}
G_{\D_+}(x_1,x_2)\sim \f{C'}{\xi(x_1,x_2)^{\left(\f{d}{2}+\sqrt{\f{d^2}{4}+m^2L^2+\f{\l}{2} \phi_{cl}^2L^2} \right)}}
\eeqa

For symmetry breaking solution we put $\phi_{cl}=0$, thus
\beqa \label{lcor1}
G_{\D_+}(x_1,x_2)\sim \f{C'}{\xi(x_1,x_2)^{\left(\f{d}{2}+\sqrt{\f{d^2}{4}+m^2L^2} \right)}}
\eeqa
The propagator shows a decay with increasing distance $\xi$.\\

\noindent
(b) \underline{Large $N$ model}

\beqa\label{ld2}
G_{\D_+}(x_1,x_2)\sim \f{C'}{ \xi(x_1,x_2)^{\left(\f{d}{2}+\sqrt{\f{d^2}{4}+M^2L^2} \right)}}
\eeqa
For symmetry breaking solution we put $M^2=0$, thus
\beqa \label{lcor2}
G_{\D_+}(x_1,x_2)\sim \f{C'}{ \xi(x_1,x_2)^{d}}
\eeqa
again observing a decay with increasing $\xi$.\\

\noindent
(2) \textbf{Neumann Boundary Condition:}\\

\noindent
(a) \underline{Single Scalar}

\beqa\label{ld3}
G_{\D_-}(x_1,x_2)\sim \f{C'}{ \xi(x_1,x_2)^{\left(\f{d}{2}-\sqrt{\f{d^2}{4}+m^2L^2+\f{\l \phi_{cl}^2L^2}{2}} \right)}}
\eeqa
For symmetry breaking solution we put $\phi_{cl}=0$, thus
\beqa \label{lcor3}
G_{\D_-}(x_1,x_2)\sim \f{C'}{ \xi(x_1,x_2)^{\left(\f{d}{2}-\sqrt{\f{d^2}{4}+m^2L^2} \right)}}
\eeqa
again we observe a decay as $\xi$ increases.\\

\noindent
(b) \underline{Large $N$ model}

\beqa\label{ld4}
G_{\D_-}(x_1,x_2)\sim \f{C'}{ \xi(x_1,x_2)^{\left(\f{d}{2}-\sqrt{\f{d^2}{4}+M^2L^2} \right)}}
\eeqa
For symmetry breaking solution we put $M^2=0$, thus
\beqa \label{lcor4}
G_\D(x_1,x_2)\sim C' 
\eeqa
In this case there exists no decay. 
\subsubsection{Behavior of correlator and symmetry breaking}
Using the results presented above  and the trace expressions computed in the previous sections we explore the possibility of symmetry breaking in various theories with the two boundary conditions by evaluating the correlator in equation (\ref{correlator}). The structure of the correlators that are presented next is the following: the last (or $\xi$ dependent) terms in the exponentials come from the large distance behaviour of the trace given in equations (\ref{ld1}), (\ref{ld2}), (\ref{ld3}) and (\ref{ld4}) while the other terms are the coincident point propagators/traces we evaluated in the previous sections.\\
\vspace{0.05cm}

\noindent
\textbf{AdS$_2$}:\\

\vspace{0.05cm}
\noindent
\underline{Large $N$, Dirichlet boundary condition:}

\beqa
\langle\phi^*(x)\phi(x_0)  \rangle &\simeq &\rho_0^2 \exp\left[\f{1}{2\pi N}\left(\psi^{(0)}\left(\sqrt{\f{1}{4}+M^2 L^2}+ \frac{1}{2}\right)+\psi^{(0)}\left(\f{1}{2}+\sqrt{\f{1}{4}+m^2 L^2}\right)\right)\right.\nonumber\\
&+&\left.\f{C}{N \xi(x_1,x_2)^{\left(\f{1}{2}+\sqrt{\f{1}{4}+M^2L^2} \right)}} \right]
\eeqa
Since the correlator in this case does not vanish when $M^2=0$ we get symmetry breaking solutions in AdS$_2$ as was seen in \cite{Kakkar:2022hub}. In contrast for flat space we have
\beqa
\langle\phi^*(x)\phi(x_0)  \rangle &\simeq &\rho_0^2 \left(\f{1}{\L|x-x_0|} \right)^{\f{1}{Nc}}
\eeqa
Here $\L$ is a $UV$ cut-off and $c$ some dimensionless constant. We observe that the correlator vanishes as $|x|\rightarrow \infty$ and thus there is no symmetry breaking in flat space.\\
\vspace{0.05cm}

\noindent
\noindent
\underline{Large $N$, Neumann boundary condition:}

\beqa
\langle\phi^*(x)\phi(x_0)  \rangle &\simeq &\rho_0^2 \exp\left[\f{1}{2\pi N}\left(\psi^{(0)}\left(\sqrt{\f{1}{4}+M^2 L^2}+ \frac{1}{2}\right)-\pi\tan\left(\pi \sqrt{\f{1}{4}+M^2 L^2}\right) \right.\right.\nonumber\\
&-&\left.\psi^{(0)}\left(\f{1}{2}+\sqrt{\f{1}{4}+m^2 L^2}\right)+\pi\tan\left(\pi \sqrt{\f{1}{4}+m^2 L^2}\right)\right)\nonumber\\
&+&\f{C}{N \xi(x_1,x_2)^{\left(\f{1}{2}-\sqrt{\f{1}{4}+M^2L^2} \right)}}\Bigg] 
\eeqa
Thus as $M^2=0$ the correlator vanishes as the first $\tan$ blows up and symmetry breaking is not possible. This reiterates our findings in section \ref{ads2largeNsection}. \\
\vspace{0.05cm}

\noindent
\underline{Single Scalar, Dirichlet boundary condition:}

\beqa
\langle\phi^*(x)\phi(x_0)  \rangle &\simeq &\rho_0^2 \exp\left[\f{1}{2\pi }\left(\psi^{(0)}\left(\f{1}{2}+\sqrt{\f{1}{4}+m^2 L^2+\f{\l}{2}\phi_{cl}^2 L^2}\right)-\psi^{(0)}\left(\f{1}{2}+\sqrt{\f{1}{4}+m^2 L^2}\right)\right) \right.\nonumber\\
&+&\left.\f{C}{ \xi(x_1,x_2)^{\left(\f{1}{2}+\sqrt{\f{1}{4}+m^2L^2+\f{\l }{2}\phi_{cl}^2L^2} \right)}}\right]
\eeqa
The correlator is finite and we get symmetry breaking solutions as seen in section \ref{ads2singles}.\\

\vspace{0.05cm}
\noindent
\underline{Single Scalar, Neumann boundary condition:}

\beqa
\langle\phi^*(x)\phi(x_0)  \rangle &\simeq &\rho_0^2 \exp\left[\f{1}{2\pi }\left(\psi^{(0)}\left(\f{1}{2}-\sqrt{\f{1}{4}+m^2 L^2+\f{\l}{2}\phi_{cl}^2 L^2}\right) -\psi^{(0)}\left(\f{1}{2}-\sqrt{\f{1}{4}+m^2 L^2}\right)\right.\right.\nonumber\\
&-&\left.\left.\pi\tan\left(\pi \sqrt{\f{1}{4}+m^2 L^2+\f{\l}{2}\phi_{cl}^2L^2} \right)+\pi\tan\left(\pi \sqrt{\f{1}{4}+m^2 L^2}\right)\right) \right.\nonumber\\
&+&\left.\f{C}{ \xi(x_1,x_2)^{\left(\f{1}{2}-\sqrt{\f{1}{4}+m^2L^2+\f{\l }{2}\phi_{cl}^2L^2} \right)}} \right]
\eeqa
Here too the correlator is finite and we get symmetry breaking solutions as seen in section \ref{ads2singles}.\\
\vspace{0.05cm}

\noindent
\textbf{AdS$_3$}:\\
\vspace{0.05cm}

\noindent
\underline{Large $N$, Dirichlet boundary condition:}

\beqa
\langle\phi^*(x)\phi(x_0)  \rangle &\simeq &\rho_0^2 \exp\left[\f{1}{N }\left(\f{(1+M^2 L^2)^{1/2}}{4 \pi }\right) +\f{C}{ N\xi(x_1,x_2)^{\left(1+\sqrt{1+M^2L^2} \right)}} \right]
\eeqa
The correlator again does not vanish as $M^2=0$ and we get symmetry breaking solutions as discussed in section \ref{ads3n}.\\
\vspace{0.05cm}

\noindent
\underline{Large $N$, Neumann boundary condition:}

\beqa
\langle\phi^*(x)\phi(x_0)  \rangle &\simeq &\rho_0^2 \exp\left[\f{1}{N }\left(-\f{(1+M^2 L^2)^{1/2}}{6 \pi}\right) +\f{C}{ N\xi(x_1,x_2)^{\left(1-\sqrt{1+M^2L^2} \right)}} \right]
\eeqa
 Thus as already seen in section \ref{ads3n} we get symmetry breaking solutions as even when $M^2=0$ the correlator does not vanish.\\
\vspace{0.05cm}

\noindent
\underline{Single Scalar, Dirichlet boundary condition:}

\beqa
\langle\phi^*(x)\phi(x_0)  \rangle \simeq \rho_0^2 \exp\left[\f{(1+M^2 L^2)^{1/2}}{4 \pi } +\f{C}{\xi(x_1,x_2)^{\left(1+\sqrt{1+m^2L^2+\f{\l }{2}\phi_{cl}^2L^2} \right)}}\right]
\eeqa
The correlator again is finite and we get symmetry breaking solution as discussed in section \ref{ads3singles} .\\
\vspace{0.05cm}

\noindent
\underline{Single Scalar, Neumann boundary condition:}

\beqa
\langle\phi^*(x)\phi(x_0)  \rangle \simeq \rho_0^2 \exp\left[-\f{(1+M^2 L^2)^{1/2}}{4 \pi}+\f{C}{\xi(x_1,x_2)^{\left(1-\sqrt{1+m^2L^2+\f{\l }{2}\phi_{cl}^2L^2} \right)}} \right]
\eeqa
This supports the occurrence of symmetry breaking solution in section \ref{ads3singles} as the correlator in this case too is finite.\\
\vspace{0.05cm}

\noindent
\textbf{AdS$_4$:}\\
\vspace{0.05cm}

\noindent
\underline{Large $N$, Dirichlet boundary condition:}

\beqa
\langle\phi^*(x)\phi(x_0)  \rangle &\simeq &\rho_0^2 \exp\left[\f{F_1}{N } +\f{C}{N \xi(x_1,x_2)^{\left(\f{3}{2}+\sqrt{\f{9}{4}+M^2L^2} \right)}}\right]
\eeqa
From equation (\ref{f1}) we can get symmetry breaking solutions as $M^2=0$ as this does not cause the correlator to vanish as was seen in \cite{Kakkar:2022hub}.
\\
\vspace{0.05cm}

\noindent
\underline{Large $N$, Neumann boundary condition:}

\beqa
\langle\phi^*(x)\phi(x_0)  \rangle &\simeq &\rho_0^2 \exp\left[\f{F_2}{N }+\f{C}{N \xi(x_1,x_2)^{\left(\f{3}{2}-\sqrt{\f{9}{4}+M^2L^2} \right)}} \right]
\eeqa
We observe from the expression of $F_2$ in \ref{f2} that the correlator vanishes as $M^2=0$ as the first $\tan$ blows up with a negative sign at this value. Thus we shall get no symmetry breaking solutions as already discussed in section \ref{ads4largeNsection}.
\\
\vspace{0.05cm}

\noindent
\underline{Single Scalar, Dirichlet  boundary condition:}

\beqa
\langle\phi^*(x)\phi(x_0)  \rangle &\simeq &\rho_0^2 \exp\left[F_3 +\f{C}{\xi(x_1,x_2)^{\left(\f{3}{2}+\sqrt{\f{9}{4}+m^2L^2+\f{\l }{2}\phi_{cl}^2L^2} \right)}}  \right]
\eeqa
Again symmetry breaking solutions are possible as the correlator does not vanish as seen from equation (\ref{f3}).\\

\vspace{0.05cm}

\noindent
\underline{Single Scalar, Neumann boundary condition:}

\beqa
\langle\phi^*(x)\phi(x_0)  \rangle &\simeq &\rho_0^2 \exp\left[F_4+\f{C}{ \xi(x_1,x_2)^{\left(\f{3}{2}-\sqrt{\f{9}{4}+m^2L^2+\f{\l }{2}\phi_{cl}^2L^2} \right)}}  \right]
\eeqa
From equation (\ref{f4}) we can get symmetry broken solutions as in this case too the correlator does not vanish thus confirming our result in section \ref{ads4singles}.
The expressions for functions $F_1 -F_4$ being quite long are given in appendix \ref{ads4corr}.

\section{Discussion}\label{discussion}

In this paper, we evaluated the partition functions for scalars in AdS$_{d+1}$ with Neumann boundary condition by analytically continuing the Dirichlet boundary condition results that was implemented by a deformation of the contour for $\l$ integration, $\l$ being the eigenvalue of the scalar Laplace operator in AdS. We then utilized these partition functions to study phases of the single scalar and $O(N)$ models with finite and large $N$ in various dimensions. The analysis is carried out by computing the effective potentials both at zero and finite temperatures as was done for the Dirichlet case in \cite{Kakkar:2022hub}. A major difference as discussed in this paper is that, the constraint $\D_{-}\ge (d-2)/2$ (which follows from the unitarity of the boundary CFT) imposes additional restrictions on the allowed regions in the phase space.  

As indicated in the introduction, for certain theories, we get unstable potentials. For the phases named unstable in Table \ref{tab1} the expectation value of the scalar field gets fixed at the end points of the potential plots. For certain others we are unable to access the symmetry breaking phase as the effective mass $(M^2)$ being equal to zero is ruled out either by functional form of the effective potentials or by the demand that the finite temperature effective potential is a convergent series. The finite temperature convergence constraint implies that even within the window allowed by unitarity we have have certain regions of the phase space being ruled out.  This thus leaves us with a few open problems.

Deformations in the dual CFT corresponding to Dirichlet boundary condition trigger renormalization group flow to Neumann boundary condition \cite{Witten:2001ua}, \cite{Berkooz:2002ug}. This has been exploited in recent works in verifying boundary F-theorems \cite{Giombi:2020rmc}, \cite{Giombi:2021cnr} through computations in AdS field theories. It is then a natural question then to ask, how the phase studies in \cite{Kakkar:2022hub} is related to the one here in this paper through such bulk/boundary deformations.

It is well known \cite{Klebanov:1999tb} that the $\D_-$ theory is not independent from the $\D_+$ theory but is, in fact, related to it by a canonical transformation that interchanges the roles of $\phi_0(x)$ and $A(x)$. Further the generating functional for correlators in the theory with one choice of boundary condition has been argued to be a Legendre transform of the generating functional in the theory with the other boundary condition. This thus leads us to the question if the microcanonical ensemble is better suited to study the theory with $\D_-$  given the instabilities arising in the canonical ensemble description. This will allow us to study the phases as a function of the energy density in place of the temperature. The latter as we have seen led to the shrinking of the allowed region within the unitarity bound. 

We hope to address these in our upcoming work.

\vspace{5.5mm}
\noindent
{\bf Acknowledgements :}
\\ 
Biki Bishwakarma acknowledges the support and financial assistance of the University Grants Commission (UGC), Government of India, through the UGC Junior Research Fellowship (JRF) scheme with Ref. No.: 231610072781. Astha Kakkar acknowledges the support and financial assistance of Council of Scientific and Industrial Research (CSIR), Government of India through the CSIR Research Associateship scheme. S.S. thanks Vidyasagar University, for the Personal Research Grant 2025-2026.

\appendix
\section{Details of thermal AdS computations}\label{thermaltrace}

In this appendix we give details of the computation of the partition function with Neumann boundary condition for thermal AdS. These calculations follow directly from \cite{Kakkar:2022hub} for Dirichlet boundary condition. We repeat them here highlighting the points of difference for the Neumann boundary condition.

\subsection{Thermal AdS$_3$}

Thermal AdS$_3$ is defined as the quotient space $\mathbb{H}^3/\mathbb{Z}$ with the metric 
\beqa
ds^2=\f{L^2}{y^2}\left(dy^2+dzd\bar{z}\right)
\eeqa
and with the following action of $\g^{n}\in \mathbb{Z}$ on the coordinates

\beqa
\g^n(y,z)=(e^{-n\beta}y, e^{2\pi in\tau}z) ~~~~\mbox{where}~~~~\tau=\frac{1}{2\pi}(\theta+i\beta)
\eeqa
In terms of real coordinates $z=x_1+ix_2$, we can write the action of the group element on the real coordinates $\vec{x}=(x_1,x_2)$ as, 

\beqa\label{realt}
\g^n \vec{x}=\left(e^{-n\beta}(x_1 \cos n\theta-x_2\sin n\theta),e^{-n\beta}(x_1 \sin n\theta+x_2\cos n\theta)\right).
\eeqa
We can now construct, using the eigenfunctions $\Psi_{\l,\vec{k}}(\vec{x},y)$ that solve (\ref{eeqn}), the following function that is invariant under the above action on the coordinates,

\beqa\label{norma}
\Phi_{\vec{k},\l}(\vec{x},y)=\frac{1}{{\cal N}}\sum_{n=-\infty}^{\infty} (ke^{-n\beta}y)K_{i\l}(k e^{-n\beta}y)e^{-i\vec{k}.(\g^n \vec{x})}.
\eeqa  

As noted in \cite{Kakkar:2022hub} it is easier to perform the integrals over the full $\mathbb{H}^3$ instead of the fundamental domain of $\mathbb{H}^3/\mathbb{Z}$. As this  results in the sum being divergent, normalization ${\cal N}$ is introduced in (\ref{norma}) to regularize the sum over $n$ as is explained further below.

We normalize  $\Phi_{\vec{k},\l}(\vec{x},y)$ in a way similar to that of the zero temperature, and setting $L=1$, as given below,

\beqa\label{normt}
&&\int d^{3}x\sqrt{g}~\Phi_{\vec{k},\l}(\vec{x},y)\Phi^{*}_{\vec{k}^{\prime},\l^{\prime}}(\vec{x},y)\\\nonumber
&=&\frac{1}{{\cal N}^2}\sum_{n,n^{\prime}} \int d^{3}x\sqrt{g}~(ke^{-n\beta}y)(k^{\prime}e^{-n^{\prime}\beta}y)K_{i\l}(k e^{-n\beta}y)K_{i\l^{\prime}}(k^{\prime} e^{-n^{\prime}\beta}y)e^{-i\vec{k}.(\g^n \vec{x})}e^{i\vec{k}^{\prime}.(\g^{n^{\prime}} \vec{x})}\\\nonumber
&=&\frac{1}{{\cal N}^2}\sum_{n,n^{\prime}} \int \frac{dy}{y}~(ke^{-n\beta})(k^{\prime}e^{-n^{\prime}\beta})K_{i\l}(k e^{-n\beta}y)K_{i\l^{\prime}}(k^{\prime} e^{-n^{\prime}\beta}y)(2\pi)^2\d^2(\g^n \vec{k}-\g^{n^{\prime}} \vec{k}^{\prime})\\\nonumber
&=&\frac{1}{{\cal N}^2}\sum_{n,n^{\prime}} \int \frac{dy}{y}~(ke^{-(n-n^{\prime})\beta})^2K_{i\l}(k e^{-n\beta}y)K_{i\l^{\prime}}(k e^{-n\beta}y)(2\pi)^2\d^2(\g^{(n-n^{\prime})} \vec{k}-\vec{k}^{\prime})\\\nonumber
&=&\frac{1}{{\cal N}}\sum_{n} (ke^{-n\b})^2(2\pi)^2\d^2(\g^{n} \vec{k}-\vec{k}^{\prime})\d(\l-\l^{\prime})/\m(\l).
\eeqa

The action of $\g^{n}$ on $\vec{k}$ is 

\[
\g^n \vec{k}=\left(e^{-n\beta}(k_1 \cos n\theta+k_2\sin n\theta),e^{-n\beta}(k_2 \cos n\theta-k_1\sin n\theta)\right)
\]

so that we have $|\g^n \vec{k}|=e^{-n\b}k$. 

In the last line, since each value of $n^{\prime}$ gives the same series in $n$, the sum over $n^{\prime}$ being divergent cancels a factor of ${\cal N}$ in the denominator. The divergence appears due to the infinite number of elements of the group $\mathbb{Z}$ (${\cal N} \rightarrow \infty$ limit of $\mathbb{Z}_{\cal N}$).  
We shall again see that we reproduce a finite expression for one-loop partition function at non-zero temperature.

The final line of (\ref{normt}) hence gives the normalization

\beqa
\frac{1}{{\cal N}}\sum_{n} \int  \frac{d^2 k}{(2\pi)^2}\f{1}{k^2} \int d\l ~\m(\l)(ke^{-n\b})^2(2\pi)^2\d^2(\g^{n} \vec{k}-\vec{k}^{\prime})\d(\l-\l^{\prime})/\m(\l)=1
\eeqa

We next compute the one-loop partition function. As in the zero temperature case, this is given by the first line of (\ref{trace1}) with the modification that, $\langle y,\vec{x}|\l,k\rangle=\Phi_{\vec{k},\l}(\vec{x},y)$. We can see that $\Phi_{\vec{k},\l}(\vec{x},y)$ have the same eigenvalues $\l$ as $\Psi_{\vec{k},\l}(\vec{x},y)$ by noting that

\beqa
\pa^2_{\vec{x}} e^{-i\vec{k}.(\g^n \vec{x})}=-|e^{-n\b}k|^2e^{-i\vec{k}.(\g^n \vec{x})}.
\eeqa
 We thus may write

\beqa
&&\tr\left[\frac{1}{- \Box_E+V^{''}(\phi_{cl})}\right]\\\nonumber &=&
\frac{1}{{\cal N}^2}\sum_{n,n^{\prime}} \int d^{3}x\sqrt{g}~\int\frac{d^2k}{(2\pi)^2}\int \f{d\l~\m(\l)}{\l^2 +\n^2}(e^{-n\beta}y)(e^{-n^{\prime}\beta}y)K_{i\l}(k e^{-n\beta}y)\nonumber\\
&\times& K_{i\l}(k e^{-n^{\prime}\beta}y)e^{-i\vec{k}.(\g^n \vec{x})}e^{i\vec{k}.(\g^{n^{\prime}}\vec{x})}\\\nonumber
&=&\frac{1}{{\cal N}^2}\sum_{n,n^{\prime}} \int \f{dy}{y}~\int\frac{d^2k}{(2\pi)^2}\int \f{d\l~\m(\l)}{\l^2 +\n^2}e^{-(n+n^{\prime})\beta} K_{i\l}(k e^{-(n-n^{\prime})\beta}y)K_{i\l}(k y)(2\pi)^2\d^2(\g^n\vec{k}-\g^{n^{\prime}}\vec{k})\\\nonumber
&=&\frac{1}{{\cal N}}\sum_{n\ne 0}\f{e^{-n\beta}}{|1-e^{2\pi i n\t}|^2} \int \f{dy}{y}~\int\frac{d^2k}{(2\pi)^2}\int \f{d\l~\m(\l)}{\l^2 +\n^2} K_{i\l}(k e^{-n\beta}y)K_{i\l}(k y)(2\pi)^2\d^2(\vec{k})
\eeqa

Each term in the summation over $n^{\prime}$ again gives the same series in $n$, thus canceling a factor of ${\cal N}$. Further, we have discarded the zero temperature $n=0$ contribution which was computed earlier. 
Next tracing the steps as in equations (\ref{ikrel})-(\ref{ikform}) and using the invariance of the expression under $n \rightarrow -n$, the trace can be written as 

\beqa
&& \frac{2}{{\cal N}}\sum_{n=1}^{\infty}\f{e^{-n\beta}}{|1-e^{2\pi i n\t}|^2} \int \f{dy}{y}~\int\frac{d^2k}{(2\pi)^2} K_{\n}(k y)I_{-\n}(k e^{-n\beta}y)(2\pi)^2\d^2(\vec{k})\\\nonumber
&=& \frac{2}{{\cal N}}\sum_{n=1}^{\infty}\int \f{dy}{y}\f{e^{-n\beta}}{|1-e^{2\pi i n\t}|^2} \f{e^{\b n\n}}{(-2\n)}
\eeqa

Defining $V''(\phi_{cl})=M^2$, so that $\n=\sqrt{1+M^2}$, the finite temperature contribution to the one-loop partition function $Z^{(1)}$ is, 

\beqa \label{intads3n}
\log Z^{(1)}_{\t}&=&\f{1}{2}\int_{M^2}^{\infty}\tr\left[\frac{1}{- \Box_E+M^2}\right] dM^2\label{paV}\\
&=& -\frac{1}{{\cal N}}\sum_{n=1}^{\infty}\int \f{dy}{y}\f{e^{-n\beta}}{|1-e^{2\pi i n\t}|^2} \int_{M^2}^{\infty}\f{e^{\b n\n}}{2\n}dM^2\\
&=&\frac{1}{{\cal N}}\sum_{n=1}^{\infty}\int \f{dy}{y}\f{1}{\b n}\f{e^{-n\beta(1-\n)}}{|1-e^{2\pi i n\t}|^2} 
\eeqa

The integral over $y$ can be written as 

\beqa
\int_0^{\infty} \f{dy}{y}=\sum_{m=-\infty}^{\infty}\int_{e^{-(m+1)\b}}^{e^{-m\b}} \f{dy}{y}
={\cal N}\b
\eeqa

where $\b$ is the value of the $y$ integral over the fundamental region 
$e^{-\b}\le y \le 1$. Finally we get the one-loop partition as,

\beqa\label{finalp3}
\log Z^{(1)}_{\t}=\sum_{n=1}^{\infty}\f{1}{n}\f{e^{-n\beta(1-\n)}}{|1-e^{2\pi i n\t}|^2} 
\eeqa

\subsection{Thermal AdS$_{d+1}$}

We now generalize the previous computation to that for higher odd dimensional spaces ($d$ even). We begin by writing down the metric
 
\beqa
ds^2=\f{L^2}{y^2}\left(dy^2+\sum_{i=1}^{d/2} dz_id\bar{z_i}\right)
\eeqa

Corresponding to each $z_i$ is associated an angular transformation $\theta_i$.
Thermal AdS$_{d+1}$ is thus the quotient space with the action of $\g^n_i$ as,

\beqa\label{realtd}
\g^n_i(y,z)=(e^{-n\beta}y, e^{2\pi in\tau_i}z_i) ~~~~\mbox{where}~~~~\tau_i=\frac{1}{2\pi}(\theta_i+i\beta)
\eeqa

Again writing, real coordinates $z_i=x_{i1}+ix_{i2}$, the action of each $\g^n_i$ on the real coordinates $\vec{x}_i=(x_{i1},x_{i2})$ can be written as in equation (\ref{realt}) with $\theta_i$.

The scalar wave function invariant under the transformation (\ref{realtd}) now is,

\beqa
\Phi_{\vec{k},\l}(\vec{x},y)=\frac{1}{{\cal N}}\sum_{n=-\infty}^{\infty} (ke^{-n\beta}y)^{d/2}K_{i\l}(k e^{-n\beta}y)e^{-i\vec{k}_i.(\g^n_i \vec{x}_i)}
\eeqa

The normalization can be worked out as follows,

\beqa\label{normt1}
&&\int d^{d+1}x\sqrt{g}~\Phi_{\vec{k},\l}(\vec{x},y)\Phi^{*}_{\vec{k}^{\prime},\l^{\prime}}(\vec{x},y)\\\nonumber
&=&\frac{1}{{\cal N}^2}\sum_{n,n^{\prime}} \int d^{d+1}x\sqrt{g}~(ke^{-n\beta}y)^{d/2}(k^{\prime}e^{-n^{\prime}\beta}y)^{d/2}K_{i\l}(k e^{-n\beta}y)K_{i\l^{\prime}}(k^{\prime} e^{-n^{\prime}\beta}y)e^{-i\vec{k}_i.(\g^n_i \vec{x}_i)}e^{i\vec{k}^{\prime}_i.(\g^{n^{\prime}}_i \vec{x}_i)}\\\nonumber
&=&\frac{1}{{\cal N}}\left[\d(\l-\l^{\prime})/\m(\l)\right]\sum_{n} \prod_{i=1}^{d/2}(ke^{-n\b})^2(2\pi)^2\d^2(\g^{n}_i \vec{k}_i-\vec{k}^{\prime}_i)
\eeqa

while the trace thus takes the following form,

\beqa
&&\tr\left[\frac{1}{- \Box_E+V^{''}(\phi_{cl})}\right]\\\nonumber &=&
\frac{1}{{\cal N}^2}\sum_{n,n^{\prime}} \int d^{3}x\sqrt{g}~\int \f{d\l~\m(\l)}{\l^2 +\n^2}\int\frac{d^dk}{(2\pi)^d}(e^{-n\beta}y)^{d/2}(e^{-n^{\prime}\beta}y)^{d/2}K_{i\l}(k e^{-n\beta}y)\nonumber\\
&\times& K_{i\l}(k e^{-n^{\prime}\beta}y)e^{-i\vec{k}_i.(\g^n_i \vec{x}_i)}e^{i\vec{k}_i.(\g^{n^{\prime}}_i\vec{x}_i)}\\\nonumber
&=&\frac{1}{{\cal N}}\sum_{n\ne 0}\int \f{dy}{y}~\int \f{d\l~\m(\l)}{\l^2 +\n^2}\int\frac{d^dk}{(2\pi)^d}  K_{i\l}(k e^{-n\beta}y)K_{i\l}(k y)\prod_{i=1}^{d/2}\f{e^{-n\beta}}{|1-e^{2\pi i n\t_i}|^2}(2\pi)^2 \d^2(\vec{k_i})\non
&=& \frac{2}{{\cal N}}\sum_{n=1}^{\infty} \int \f{dy}{y}~\int\frac{d^dk}{(2\pi)^d} K_{\n}(k y)I_{-\n}(k e^{-n\beta}y)\prod_{i=1}^{d/2}\f{e^{-n\beta}}{|1-e^{2\pi i n\t_i}|^2}(2\pi)^2\d^2(\vec{k_i})\\\nonumber
&=& \frac{2}{{\cal N}}\sum_{n=1}^{\infty}\int \f{dy}{y} \f{e^{\b n\n}}{-2\n}
\prod_{i=1}^{d/2}\f{e^{-n\beta}}{|1-e^{2\pi i n\t_i}|^2}
\eeqa

This gives us the following expression for the partition function

\beqa\label{zd}
\log Z=\sum_{n=1}^{\infty}\f{e^{n\beta\n}}{n}\prod_{i=1}^{d/2}\f{e^{-n\beta}}{|1-e^{2\pi i n\t_i}|^2} 
\eeqa

For $\theta_i=0$ equation (\ref{zd}) reduces to

\beqa\label{zdt0}
\log Z=\sum_{n=1}^{\infty}\f{1}{n}\f{e^{-n\beta(d/2-\n)}}{|1-e^{-n\beta}|^d} 
\eeqa

For when $d$ is odd we write $z_i=x_{i1}+ix_{i2}$ with $i=1,\cdots (d-1)/2$ and the metric

\beqa
ds^2=\f{L^2}{y^2}\left(dy^2+\sum_{i=1}^{(d-1)/2} dz_id\bar{z_i}+dx_d^2\right)
\eeqa

while $\g^n$ acts on the coordinates as

\beqa
\g^n_i(y,z_i,x_{d})=(e^{-n\beta}y, e^{2\pi in\tau_i}z_i,e^{-n\beta}x_d).
\eeqa  

The $\g^n$ invariant solution now is, 

\beqa
\Phi_{\vec{k},\l}(\vec{x},y)=\frac{1}{{\cal N}}\sum_{n=-\infty}^{\infty} (ke^{-n\beta}y)^{d/2}K_{i\l}(k e^{-n\beta}y)e^{-i\vec{k}_i.(\g^n_i \vec{x}_i)}e^{-i{k}_d(e^{-n\b}{x}_d)}
\eeqa  

which can be normalized as

\beqa\label{normtdo}
&&\int d^{d+1}x\sqrt{g}~\Phi_{\vec{k},\l}(\vec{x},y)\Phi^{*}_{\vec{k}^{\prime},\l^{\prime}}(\vec{x},y)\\\nonumber
&=&\frac{1}{{\cal N}}\left[\d(\l-\l^{\prime})/\m(\l)\right]\sum_{n} (ke^{-n\b})(2\pi)\d(e^{-n\b}{k}_d-{k}^{\prime}_d)\prod_{i=1}^{(d-1)/2}(ke^{-n\b})^2(2\pi)^2\d^2(\g^{n}_i \vec{k}_i-\vec{k}^{\prime}_i)
\eeqa

The partition function in this case becomes

\beqa\label{zdo}
\log Z=\sum_{n=1}^{\infty}\f{e^{-n\beta(1/2-\n)}}{n|1-e^{-n\beta}|}\prod_{i=1}^{(d-1)/2}\f{e^{-n\beta}}{|1-e^{2\pi i n\t_i}|^2} 
\eeqa

For $\th_i=0$, the odd $d$ case also leads to the expression (\ref{zdt0}) for the one-loop partition function. As noted before the appearance of $\D_{-}=d/2-\n$ in (\ref{zd}), (\ref{zdo}) is due to the choice of our contour in Figure \ref{contour2}.

\section{Expressions for functions appearing in AdS$_4$ correlators}\label{ads4corr}

\beqa\label{f1}
F_1&=&-\f{(2+M^2 L^2)}{32\pi^2}\left(\psi\left(\sqrt{\f{9}{4}+M^2 L^2}-\f{1}{2} \right) +\psi\left(\sqrt{\f{9}{4}+M^2 L^2}+\f{3}{2} \right)\right.\nonumber\\
&-&\left.\psi\left(\sqrt{\f{9}{4}+m^2 L^2}-\f{1}{2} \right) -\psi\left(\sqrt{\f{9}{4}+m^2 L^2}+\f{3}{2} \right)\right)
\eeqa

\beqa \label{f2}
F_2&=&-\f{\l(2+M^2 L^2)}{32 \pi^2}\left[ \psi\left(\f{3}{2}+\sqrt{\f{9}{4}+M^2 L^2}\right)+\psi\left(-\f{1}{2}+\sqrt{\f{9}{4}+M^2 L^2}\right)\right.\nonumber\\
&-&\psi\left(\f{3}{2}+\sqrt{\f{9}{4}+m^2 L^2}\right)-\psi\left(-\f{1}{2}+\sqrt{\f{9}{4}+m^2 L^2}\right)-2\pi \tan\left(\pi \sqrt{\f{9}{4}+M^2 L^2}\right)\nonumber\\
&+&\left.2\pi\tan\left(\pi \sqrt{\f{9}{4}+m^2 L^2}\right) \right] 
\eeqa

\beqa\label{f3}
F_3&=&-\f{(2+M^2 L^2)}{32 \pi^2}\left[ \psi\left(\f{3}{2}+\sqrt{\f{9}{4}+m^2L^2+\f{\l}{2}\phi_{cl}^2 L^2}\right)+\psi\left(-\f{1}{2}+\sqrt{\f{9}{4}+m^2L^2+\f{\l}{2}\phi_{cl}^2 L^2}\right)\right.\nonumber\\
&-&\left.\psi\left(\f{3}{2}+\sqrt{\f{9}{4}+m^2L^2}\right)-\psi\left(-\f{1}{2}+\sqrt{\f{9}{4}+m^2L^2}\right) \right]
\eeqa

\beqa\label{f4}
F_4&=&-\f{\l(2+M^2 L^2)}{32 \pi^2}\left[ \psi\left(\f{3}{2}+\sqrt{\f{9}{4}+m^2L^2+\f{\l}{2}\phi_{cl}^2 L^2}\right)+\psi\left(-\f{1}{2}+\sqrt{\f{9}{4}+m^2L^2+\f{\l}{2}\phi_{cl}^2 L^2}\right)\right.\nonumber\\
&-&\psi\left(\f{3}{2}+\sqrt{\f{9}{4}+m^2L^2}\right)-\psi\left(-\f{1}{2}+\sqrt{\f{9}{4}+m^2L^2}\right)-2\pi \tan\left(\pi \sqrt{\f{9}{4}+m^2L^2+\f{\l}{2}\phi_{cl}^2 L^2}\right)\nonumber\\
&+&\left.2\pi\tan\left(\pi \sqrt{\f{9}{4}+m^2L^2}\right) \right]
\eeqa


\begin{thebibliography}{99}


\bibitem{Callan:1989em}
C.~G.~Callan, Jr. and F.~Wilczek,
``INFRARED BEHAVIOR AT NEGATIVE CURVATURE,''
Nucl. Phys. B \textbf{340} (1990), 366-386
doi:10.1016/0550-3213(90)90451-I


\bibitem{Inami:1985dj}
T.~Inami and H.~Ooguri,
``NAMBU-GOLDSTONE BOSONS IN CURVED SPACE-TIME,''
Phys. Lett. B \textbf{163} (1985), 101-105
doi:10.1016/0370-2693(85)90201-1


\bibitem{Burgess:1984ti}
C.~P.~Burgess and C.~A.~Lutken,
``Propagators and Effective Potentials in Anti-de Sitter Space,''
Phys. Lett. B \textbf{153} (1985), 137-141
doi:10.1016/0370-2693(85)91415-7



\bibitem{Inami:1985wu}
T.~Inami and H.~Ooguri,
``One Loop Effective Potential in Anti-de Sitter Space,''
Prog. Theor. Phys. \textbf{73} (1985), 1051
doi:10.1143/PTP.73.1051



\bibitem{Kamela:1998mb}
M.~Kamela and C.~P.~Burgess,
``Massive scalar effective actions on Anti-de Sitter space-time,''
Can. J. Phys. \textbf{77} (1999), 85-99
doi:10.1139/cjp-77-2-85
[arXiv:hep-th/9808107 [hep-th]].

\bibitem{Camporesi:1990wm}
R.~Camporesi,
``Harmonic analysis and propagators on homogeneous spaces,''
Phys. Rept. \textbf{196} (1990), 1-134
doi:10.1016/0370-1573(90)90120-Q

\bibitem{Camporesi:1991nw}
R.~Camporesi,
``zeta function regularization of one loop effective potentials in anti-de Sitter space-time,''
Phys. Rev. D \textbf{43} (1991), 3958-3965
doi:10.1103/PhysRevD.43.3958

\bibitem{Bytsenko:1994bc}
A.~A.~Bytsenko, G.~Cognola, L.~Vanzo and S.~Zerbini,
``Quantum fields and extended objects in space-times with constant curvature spatial section,''
Phys. Rept. \textbf{266} (1996), 1-126
doi:10.1016/0370-1573(95)00053-4
[arXiv:hep-th/9505061 [hep-th]].

\bibitem{Caldarelli:1998wk}
M.~M.~Caldarelli,
``Quantum scalar fields on anti-de Sitter space-time,''
Nucl. Phys. B \textbf{549} (1999), 499-515
doi:10.1016/S0550-3213(99)00137-6
[arXiv:hep-th/9809144 [hep-th]].


\bibitem{Gubser:2002zh}
S.~S.~Gubser and I.~Mitra,
``Double trace operators and one loop vacuum energy in AdS / CFT,''
Phys. Rev. D \textbf{67} (2003), 064018
doi:10.1103/PhysRevD.67.064018
[arXiv:hep-th/0210093 [hep-th]].

\bibitem{Das:2006wg}
A.~K.~Das and G.~V.~Dunne,
``Large-order Perturbation Theory and de Sitter/Anti de Sitter Effective Actions,''
Phys. Rev. D \textbf{74} (2006), 044029
doi:10.1103/PhysRevD.74.044029
[arXiv:hep-th/0607168 [hep-th]].



\bibitem{Aharony:2010ay}
O.~Aharony, D.~Marolf and M.~Rangamani,
``Conformal field theories in anti-de Sitter space,''
JHEP \textbf{02} (2011), 041
doi:10.1007/JHEP02(2011)041
[arXiv:1011.6144 [hep-th]].





\bibitem{Aharony:2012jf}
O.~Aharony, M.~Berkooz, D.~Tong and S.~Yankielowicz,
``Confinement in Anti-de Sitter Space,''
JHEP \textbf{02} (2013), 076
doi:10.1007/JHEP02(2013)076
[arXiv:1210.5195 [hep-th]].

\bibitem{Miyagawa:2015sql}
T.~Miyagawa, N.~Shiba and T.~Takayanagi,
``Double-Trace Deformations and Entanglement Entropy in AdS,''
Fortsch. Phys. \textbf{64} (2016), 92-105
doi:10.1002/prop.201500098
[arXiv:1511.07194 [hep-th]].

\bibitem{Sugishita:2016iel}
S.~Sugishita, ``Entanglement entropy for free scalar fields in AdS,''
JHEP \textbf{09} (2016), 128
doi:10.1007/JHEP09(2016)128
[arXiv:1608.00305 [hep-th]].



\bibitem{Carmi:2018qzm}
D.~Carmi, L.~Di Pietro and S.~Komatsu,
``A Study of Quantum Field Theories in AdS at Finite Coupling,''
JHEP \textbf{01} (2019), 200
doi:10.1007/JHEP01(2019)200
[arXiv:1810.04185 [hep-th]].

\bibitem{Kruczenski:2022lot}
M.~Kruczenski, J.~Penedones and B.~C.~van Rees,
``Snowmass White Paper: S-matrix Bootstrap,''
[arXiv:2203.02421 [hep-th]].




\bibitem{Ankur:2023lum}
Ankur, D.~Carmi and L.~Di Pietro,
``Scalar QED in AdS,''
JHEP \textbf{10} (2023), 089
doi:10.1007/JHEP10(2023)089
[arXiv:2306.05551 [hep-th]].

\bibitem{Giombi:2020rmc}
S.~Giombi and H.~Khanchandani,
``CFT in AdS and boundary RG flows,''
JHEP \textbf{11} (2020), 118
doi:10.1007/JHEP11(2020)118
[arXiv:2007.04955 [hep-th]].

\bibitem{Giombi:2021cnr}
S.~Giombi, E.~Helfenberger and H.~Khanchandani,
``Fermions in AdS and Gross-Neveu BCFT,''
JHEP \textbf{07} (2022), 018
doi:10.1007/JHEP07(2022)018
[arXiv:2110.04268 [hep-th]].


\bibitem{Nishioka:2021uef}
T.~Nishioka and Y.~Sato,
``Free energy and defect $C$-theorem in free scalar theory,''
JHEP \textbf{05} (2021), 074
doi:10.1007/JHEP05(2021)074
[arXiv:2101.02399 [hep-th]].

\bibitem{Carmi:2019ocp}
D.~Carmi,
``Loops in AdS: From the Spectral Representation to Position Space,''
JHEP \textbf{06} (2020), 049
doi:10.1007/JHEP06(2020)049
[arXiv:1910.14340 [hep-th]].

\bibitem{Carmi:2021dsn}
D.~Carmi,
``Loops in AdS: from the spectral representation to position space. Part II,''
JHEP \textbf{07} (2021), 186
doi:10.1007/JHEP07(2021)186
[arXiv:2104.10500 [hep-th]].


\bibitem{Copetti:2023sya}
C.~Copetti, L.~Di Pietro, Z.~Ji and S.~Komatsu,
``Taming Mass Gaps with Anti{\textendash}de Sitter Space,''
Phys. Rev. Lett. \textbf{133} (2024) no.8, 081601
doi:10.1103/PhysRevLett.133.081601
[arXiv:2312.09277 [hep-th]].




\bibitem{Ciccone:2024guw}
R.~Ciccone, F.~De Cesare, L.~Di Pietro and M.~Serone,
``Exploring confinement in Anti-de Sitter space,''
JHEP \textbf{12} (2024), 218
[erratum: JHEP \textbf{06} (2025), 037]
doi:10.1007/JHEP12(2024)218
[arXiv:2407.06268 [hep-th]].


\bibitem{DiPietro:2025ozw}
L.~Di Pietro, S.~R.~Kousvos, M.~Meineri, A.~Piazza, M.~Serone and A.~Vichi,
``A Bootstrap Study of Confinement in AdS,''
[arXiv:2512.00150 [hep-th]].



\bibitem{Ciccone:2025dqx}
R.~Ciccone, F.~De Cesare, L.~Di Pietro and M.~Serone,
``QCD in AdS,''
JHEP \textbf{04} (2026), 130
doi:10.1007/JHEP04(2026)130
[arXiv:2511.04752 [hep-th]].

\bibitem{Bason:2025sxb}
D.~Bason, C.~Copetti, L.~Di Pietro, Z.~Ji and S.~Komatsu,
``F-theorem for Quantum Field Theories in Anti-de Sitter Space,''
[arXiv:2512.18392 [hep-th]].

\bibitem{Bason:2025zpy}
D.~Bason, C.~Copetti, L.~Di Pietro and Z.~Ji,
``$ \mathcal{N}=2 $ super Yang-Mills in AdS$_{4}$ and F$_{AdS}$-maximization,''
JHEP \textbf{03} (2026), 254
doi:10.1007/JHEP03(2026)254
[arXiv:2506.05162 [hep-th]].







\bibitem{Ankur:2026ylr}
Ankur, L.~Di Pietro, V.~Gorbenko, S.~Komatsu and V.~Sacchi,
``Dressing and Screening in Anti-de Sitter,''
[arXiv:2601.04321 [hep-th]].


\bibitem{Antunes:2025iaw}
A.~Antunes, N.~Levine and M.~Meineri,
``Demystifying integrable QFTs in AdS: No-go theorems for higher-spin charges,''
SciPost Phys. \textbf{20} (2026), 088
doi:10.21468/SciPostPhys.20.3.088
[arXiv:2502.06937 [hep-th]].

\bibitem{Meineri:2023mps}
M.~Meineri, J.~Penedones and T.~Spirig,
``Renormalization group flows in AdS and the bootstrap program,''
[arXiv:2305.11209 [hep-th]].

\bibitem{Lauria:2023uca}
E.~Lauria, M.~Milam and B.~C.~van Rees,
``Perturbative RG flows in AdS: an \'etude,''
[arXiv:2309.10031 [hep-th]].




\bibitem{Antunes:2021abs}
A.~Antunes, M.~S.~Costa, J.~Penedones, A.~Salgarkar and B.~C.~van Rees,
``Towards bootstrapping RG flows: sine-Gordon in AdS,''
JHEP \textbf{12} (2021), 094
doi:10.1007/JHEP12(2021)094
[arXiv:2109.13261 [hep-th]].


\bibitem{Antunes:2024hrt}
A.~Antunes, E.~Lauria and B.~C.~van Rees,
``A bootstrap study of minimal model deformations,''
JHEP \textbf{05} (2024), 027
doi:10.1007/JHEP05(2024)027
[arXiv:2401.06818 [hep-th]].



\bibitem{Breitenlohner:1982jf}
P.~Breitenlohner and D.~Z.~Freedman,
``Stability in Gauged Extended Supergravity,''
Annals Phys. \textbf{144} (1982), 249
doi:10.1016/0003-4916(82)90116-6


\bibitem{Aharony:1999ti}
O.~Aharony, S.~S.~Gubser, J.~M.~Maldacena, H.~Ooguri and Y.~Oz,
``Large N field theories, string theory and gravity,''
Phys. Rept. \textbf{323} (2000), 183-386
doi:10.1016/S0370-1573(99)00083-6
[arXiv:hep-th/9905111 [hep-th]].

\bibitem{Klebanov:1999tb}
I.~R.~Klebanov and E.~Witten,
``AdS / CFT correspondence and symmetry breaking,''
Nucl. Phys. B \textbf{556} (1999), 89-114
doi:10.1016/S0550-3213(99)00387-9
[arXiv:hep-th/9905104 [hep-th]].



\bibitem{Krishnan:2016mcj}
C.~Krishnan and A.~Raju,
``A Neumann Boundary Term for Gravity,''
Mod. Phys. Lett. A \textbf{32} (2017) no.14, 1750077
doi:10.1142/S0217732317500778
[arXiv:1605.01603 [hep-th]].

\bibitem{Krishnan:2016dgy}
C.~Krishnan, A.~Raju and P.~N.~Bala Subramanian,
``Dynamical boundary for anti{\textendash}de Sitter space,''
Phys. Rev. D \textbf{94} (2016) no.12, 126011
doi:10.1103/PhysRevD.94.126011
[arXiv:1609.06300 [hep-th]].


\bibitem{Gibbons:2006ij}
G.~W.~Gibbons, M.~J.~Perry and C.~N.~Pope,
``Partition functions, the Bekenstein bound and temperature inversion in anti-de Sitter space and its conformal boundary,''
Phys. Rev. D \textbf{74} (2006), 084009
doi:10.1103/PhysRevD.74.084009
[arXiv:hep-th/0606186 [hep-th]].

\bibitem{Giombi:2008vd}
S.~Giombi, A.~Maloney and X.~Yin,
``One-loop Partition Functions of 3D Gravity,''
JHEP \textbf{08} (2008), 007
doi:10.1088/1126-6708/2008/08/007
[arXiv:0804.1773 [hep-th]].


\bibitem{Denef:2009kn}
F.~Denef, S.~A.~Hartnoll and S.~Sachdev,
``Black hole determinants and quasinormal modes,''
Class. Quant. Grav. \textbf{27} (2010), 125001
doi:10.1088/0264-9381/27/12/125001
[arXiv:0908.2657 [hep-th]].


\bibitem{David:2009xg}
J.~R.~David, M.~R.~Gaberdiel and R.~Gopakumar,
``The Heat Kernel on AdS(3) and its Applications,''
JHEP \textbf{04} (2010), 125
doi:10.1007/JHEP04(2010)125
[arXiv:0911.5085 [hep-th]].

\bibitem{Gopakumar:2011qs}
R.~Gopakumar, R.~K.~Gupta and S.~Lal,
``The Heat Kernel on $AdS$,''
JHEP \textbf{11} (2011), 010
doi:10.1007/JHEP11(2011)010
[arXiv:1103.3627 [hep-th]].


\bibitem{Gupta:2012he}
R.~K.~Gupta and S.~Lal,
``Partition Functions for Higher-Spin theories in AdS,''
JHEP \textbf{07} (2012), 071
doi:10.1007/JHEP07(2012)071
[arXiv:1205.1130 [hep-th]].

\bibitem{Keeler:2014hba}
C.~Keeler and G.~S.~Ng,
``Partition Functions in Even Dimensional AdS via Quasinormal Mode Methods,''
JHEP \textbf{06} (2014), 099
doi:10.1007/JHEP06(2014)099
[arXiv:1401.7016 [hep-th]].


\bibitem{Martin:2019flv}
V.~L.~Martin and A.~Svesko,
``Normal modes in thermal AdS via the Selberg zeta function,''
SciPost Phys. \textbf{9} (2020), 009
doi:10.21468/SciPostPhys.9.1.009
[arXiv:1910.11913 [hep-th]].

\bibitem{Kraus:2020nga}
P.~Kraus, S.~Megas and A.~Sivaramakrishnan,
``Anomalous dimensions from thermal AdS partition functions,''
JHEP \textbf{10} (2020), 149
doi:10.1007/JHEP10(2020)149
[arXiv:2004.08635 [hep-th]].


\bibitem{Kakkar:2022hub}
A.~Kakkar and S.~Sarkar,
``On partition functions and phases of scalars in AdS,''
JHEP \textbf{07} (2022), 089
doi:10.1007/JHEP07(2022)089
[arXiv:2201.09043 [hep-th]].

\bibitem{Kakkar:2023gzu}
A.~Kakkar and S.~Sarkar,
``Phases of theories with fermions in AdS,''
JHEP \textbf{06} (2023), 009
doi:10.1007/JHEP06(2023)009
[arXiv:2303.02711 [hep-th]].

\bibitem{Kakkar:2023ijc}
A.~Kakkar and S.~Sarkar,
``Partition functions for U(1) vectors and phases of scalar QED in AdS,''
JHEP \textbf{06} (2024), 095
doi:10.1007/JHEP06(2024)095
[arXiv:2311.06045 [hep-th]].

\bibitem{Kakkar:2024sso}
A.~Kakkar and S.~Sarkar,
``One-Loop Analysis of~Phases of~Scalar Field Theories in~Thermal Anti-de Sitter Spaces,''
Springer Proc. Phys. \textbf{304} (2024), 52-56
doi:10.1007/978-981-97-0289-3{\_}10



\bibitem{Kakkar:2026nqp}
A.~Kakkar and S.~Sarkar,
``Partition Functions and~Phases of~Quantum Field Theories in~AdS Spaces,''
Springer Proc. Phys. \textbf{432} (2026), 499-502
doi:10.1007/978-981-95-1513-4{\_}114










\bibitem{Mermin:1966fe}
N.~D.~Mermin and H.~Wagner,
``Absence of ferromagnetism or antiferromagnetism in one-dimensional or two-dimensional isotropic Heisenberg models,''
Phys. Rev. Lett. \textbf{17} (1966), 1133-1136
doi:10.1103/PhysRevLett.17.1133. 

\bibitem{Coleman:1973ci}
S.~R.~Coleman,
``There are no Goldstone bosons in two-dimensions,''
Commun. Math. Phys. \textbf{31} (1973), 259-264
doi:10.1007/BF01646487




\bibitem{Witten:2001ua}
E.~Witten,
``Multitrace operators, boundary conditions, and AdS / CFT correspondence,''
[arXiv:hep-th/0112258 [hep-th]].




\bibitem{Berkooz:2002ug}
M.~Berkooz, A.~Sever and A.~Shomer,
``'Double trace' deformations, boundary conditions and space-time singularities,''
JHEP \textbf{05} (2002), 034
doi:10.1088/1126-6708/2002/05/034
[arXiv:hep-th/0112264 [hep-th]].






\bibitem{Grad}
I.S. Gradshteyn and I.M. Ryzhik,
``Table of Integrals, Series, and Products".

\bibitem{Watson}
G. N. Watson
``A treatise on the theory of Bessel functions".

\end{thebibliography}
\end{document}